\documentclass[conference]{IEEEtran}
\IEEEoverridecommandlockouts

\usepackage{cite}
\usepackage{amsmath,amssymb,amsfonts}
\usepackage{algorithmic}
\usepackage{graphicx}
\usepackage{textcomp}
\usepackage[table,xcdraw]{xcolor}
\usepackage{tikz}
\usepackage{soul}
\usepackage{version}
\usepackage{subcaption}
\usepackage{makecell}
\usepackage[most]{tcolorbox}
\usepackage{pifont}
\usepackage{tabularx}
\usepackage{xspace}
\usepackage{url}

\includeversion{extended}
\excludeversion{conference}

\newcommand{\tako}{t\"{a}k\={o}}
\newcommand{\papertitle}[1]{t\"{a}k\={o}Formal}

\newcommand{\ext}[1]{\begin{extended}#1\end{extended}}
\newcommand{\conf}[1]{\begin{conference}#1\end{conference}}

\newcommand{\onmiss}{\texttt{OnMiss}}
\newcommand{\onevict}{\texttt{OnEvict}}
\newcommand{\onwb}{\texttt{OnWB}}
\newcommand{\ism}{ISM\xspace}

\definecolor{lightred}{HTML}{F8CECC}
\definecolor{darkerred}{HTML}{B85450}
\definecolor{nopColour}{HTML}{BAC8D3}
\definecolor{nopColourborder}{HTML}{23445D}

\newcommand*\circled[1]{\protect\tikz[baseline=(char.base)]{
\protect\node[shape=circle,fill=lightred,draw=darkerred,inner sep=1pt] (char) {#1};}}

\def\BibTeX{{\rm B\kern-.05em{\sc i\kern-.025em b}\kern-.08em
    T\kern-.1667em\lower.7ex\hbox{E}\kern-.125emX}}
\begin{document}

\title{\papertitle{}: Enabling Robust Software for Programmable Memory Hierarchies\ext{\\ (Extended Version)}
}

\author{Pranav Srinivasan\qquad Manos Kapritsos\qquad Yatin A. Manerkar\\\vspace{2pt}
University of Michigan\\\vspace{2pt}
\{pransrin, manosk, manerkar\}@umich.edu}

\maketitle

\begin{abstract}

Accelerators provide large performance and energy-efficiency benefits, but can significantly change the hardware-software interface.
The \tako{} programmable memory hierarchy accelerates data movement by enabling programmers to run user-defined callback functions triggered by cache misses, evictions, and writebacks. However, it also leads to drastically increased complexity and counterintuitive outcomes.
In response, we develop an ISA-level memory consistency model (MCM) for \tako{} that captures the semantics of its operation, and we show how it enables programmers to formally reason about their \tako{} programs.
We also prove the soundness of this ISA-level MCM by constructing a detailed \tako{} implementation model and verifying that all executions of the implementation model are allowed by our ISA-level MCM.
Along the way, we discover useful insights about microarchitectural modeling and verification that are applicable to hardware in general. 

\ext{This is the extended version of the ISCA 2026 paper ``\papertitle{}: Enabling Robust Software for Programmable Memory Hierarchies''. This version adds material on additional litmus tests to Section~\ref{sec:tako_litmus} to further explore the programmability of \tako{} using our ISA-level MCM.}

\end{abstract}
\section{Introduction}
\label{sec:intro}

With the end of Moore’s Law, hardware innovation has moved towards increasing performance via accelerator-level parallelism\cite{hill_accelerator-level_2021}.
Innovation in this direction involves hardware and software changes that provide significant performance and energy-efficiency gains.

These benefits come with new challenges. Accelerators today have various shapes and sizes, and often change the hardware-software interface~\cite{jouppi_tpu_2023,schwedock_tako_2022, van_der_hagen_client-optimized_2022}.
These changes make it difficult for non-experts to understand, program, and verify such systems. Precisely defining a hardware-software interface and verifying implementations against it is essential. In the past, a lack of careful reasoning about this interface has led to critical vulnerabilities, unintuitive program outcomes, and design specification ambiguities~\cite{arm_cortex-a9_2011,kocher_spectre_2019,trippel_tricheck_2017,manerkar_counterexamples_2016,lahav_repairing_2017}.

Formal methods have been used to define hardware-software interface specifications for traditional architectures for many years~\cite{berghofer_better_2009,sarkar_understanding_2011,alglave_herding_2014,pulte_simplifying_2018}, as well as to verify hardware implementations against such specifications~\cite{burch_automatic_1994,manerkar_rtlcheck_2017,choi_kami_2017}.
However, barring a few exceptions (e.g.,~\cite{subramanyan_template-based_2015,ambal_semantics_2024,tan_formalising_2025}), most formal methods work still assumes a traditional view of the computing stack rather than the accelerator-rich landscape of today.
Furthermore, the accelerator design space is so rich and varied that one cannot create a single effective methodology for formally specifying and verifying all possible accelerators.
Still, architects and formal methods experts must work together to develop new techniques for modeling and verifying classes of accelerators that have not been previously studied.

\begin{figure}[t]
  \centering

  \includegraphics{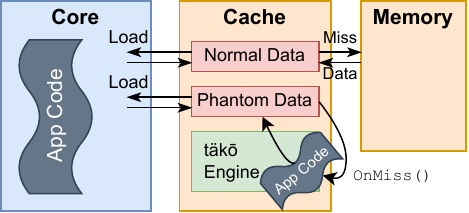}

  \caption{Image and caption from \cite{schwedock_tako_2022} showing the organization of a \tako{} program. An application registers an address range whose semantics are defined by software callbacks. These callbacks run in-cache on programmable engines.}
  \label{fig:tako_flow}
  \vspace{-15pt}
\end{figure}

In this work, we focus on developing a verified formal hardware-software interface for the \tako{}~\cite{schwedock_tako_2022} programmable memory hierarchy (PMH).
Figure~\ref{fig:tako_flow} shows \tako{}'s high-level operation: users can write \emph{callbacks} that run on cache misses, evictions, and writebacks. These callbacks give the user increased control over data movement, enabling various performance and energy-efficiency benefits.

We chose to formally model and verify \tako{} for multiple reasons.
Firstly, \tako{} fundamentally changes the hardware-software interface. In addition to its callbacks triggered by cache events, it supports \emph{phantom addresses} which are not backed by main memory. These features lead to complicated executions and counterintuitive program outcomes, requiring programmers to understand the intricacies of the memory hierarchy (e.g., details of prefetching and replacement policies) to understand how their programs will behave.
This complexity makes \tako{} a challenging and worthwhile case study for formal methods.
Secondly, \tako{} is intended to be a general-purpose accelerator. The \tako{} paper shows how \tako{} can be used to improve the performance and energy efficiency of a diverse set of workloads, including graph traversals, scatter-updates, and non-volatile memory transactions. Thus, insights gained from modeling and verifying \tako{} are likely to be more broadly applicable than those gained from modeling a more specialized accelerator.
Finally, no implementation of \tako{} currently exists beyond the closed-source simulator used for the \tako{} paper.
Thus, there is no way for researchers other than the authors of \tako{} to validate whether a \tako{} implementation is correct.
A formal hardware-software interface for \tako{} would enable such verification, and we create such an interface in this paper.

\begin{figure}
  \centering
    \begin{subfigure}{\linewidth}
        \centering
        \begin{tabular}{|ll|}
            \hline
            \multicolumn{1}{|c|}{Core 0}                       & \multicolumn{1}{c|}{{[}x{]}.\onmiss{}}       \\ \hline
            \multicolumn{1}{|l|}{(i1) [x] $\leftarrow$ 1}  & \multicolumn{1}{l|}{(i3) [x] $\leftarrow$ 2}  \\
            \multicolumn{1}{|l|}{(i2) r1 $\leftarrow$ {[}x{]}} &                             \\ \hline
        \end{tabular}
        \caption{}
        \vspace{2pt}
    \end{subfigure}

    \begin{subfigure}{\linewidth}
        \centering
        \begin{tabular}{|ll|}
            \hline
            \multicolumn{1}{|c|}{Core 0}              & \multicolumn{1}{c|}{[x].\onmiss{}}  \\ \hline
            \multicolumn{1}{|l|}{(1) \color{red}[x] misses in cache} & \multicolumn{1}{l|}{}                           \\
            \multicolumn{1}{|l|}{}                    & \multicolumn{1}{l|}{(2) (i3$)_1$ [x] $\leftarrow$ 2 }  \\
            \multicolumn{1}{|l|}{(3) (i1) [x] $\leftarrow$ 1}       & \multicolumn{1}{l|}{}                             \\
            \multicolumn{1}{|l|}{(4) \color{red} [x] evicted}         & \multicolumn{1}{l|}{}                             \\
            \multicolumn{1}{|l|}{(5) \color{red} [x] misses in cache} & \multicolumn{1}{l|}{}                            \\
            \multicolumn{1}{|l|}{}                    & \multicolumn{1}{l|}{(6) (i3$)_2$ [x] $\leftarrow$ 2}                    \\
            \multicolumn{1}{|l|}{(7) (i2) r1 $\leftarrow$ [x]}       & \multicolumn{1}{l|}{}                             \\ \hline
        \end{tabular}
        \caption{}
    \end{subfigure}
      \caption{(a) A sample \tako{} program. (b) A possible execution of said program. The evictions and misses from the cache, which were previously hidden hardware details, now impact the outcome of the program.}
  \label{fig:sample_exec}
  \vspace{-15pt}
\end{figure}

As an example of how \tako{} programs can execute in counterintuitive ways, consider the program in Figure \ref{fig:sample_exec}a, in which a program thread writes to an address [x] and subsequently reads from it.
In this case, [x] is a phantom address with an \onmiss{} callback registered for it. When an access to [x] misses in the cache, the callback runs, populating the cache with a value of 2 for [x]. In an execution where [x] is brought into the cache to execute (i1) and remains there for the execution of (i2), the value of 2 would be overwritten to 1 by (i1), and thus (i2) would read the value of 1 into r1. However, if an eviction occurs between these instructions, (i2) would miss in the cache, causing the \onmiss{} to be invoked again. In this case, the previously written value is \textit{dropped entirely}, since phantom addresses are not backed by main memory. An execution illustrating this counterintuitive behavior is shown in Figure \ref{fig:sample_exec}b. In this case, the intervening cache eviction and subsequent \onmiss{} cause the value that is loaded by (i2) to completely forget the occurrence of the previous write.
In such a system, the previously hidden details of cache features, such as a prefetching or cache replacement policy, now have a direct impact on the functional results of the program.

\tako{}'s linkage of cache features to program results fundamentally changes the memory consistency model (MCM) of an ISA that may implement \tako{}.
MCMs constrain the values that can be read by load instructions in parallel programs, so precisely specifying MCMs and verifying their implementations is critical to parallel system correctness.
A formally specified MCM for an architecture also enables proving correctness of compilation to that architecture, as well as program synthesis~\cite{gulwani:program} (code generation with correctness guarantees) for that architecture.
Defining the MCM of an architecture like \tako{} requires reasoning beyond what is used in traditional systems, because conventional MCMs have no notion of phantom addresses or cache-event-triggered callbacks.

In this work, we develop new formalisms for reasoning about cache events, callbacks, and phantom addresses to create a new ISA-level MCM for \tako{} (\S\ref{sec:axiomatic}).
This MCM is \emph{axiomatic}, i.e., executions must obey a set of axioms (properties) to be correct under the MCM.
In \S\ref{sec:tako_litmus}, we show how programmers can use our MCM to reason about realistic \tako{} programs. \conf{We provide additional litmus tests covering multiple other cases of our model in our extended version~\cite{srinivasan_takoformal_extended_2026}.}

To verify that our MCM accurately captures \tako{} functionality, we create a detailed \emph{operational} (state machine-based) model of a \tako{} implementation (\S\ref{sec:operational}).
We then formally prove (\S\ref{sec:bridge}) that for all programs, any execution possible on the operational model is also allowed by our ISA-level MCM.
This proof is machine-checked, which means that a verification engine ensures that the steps we write in our proof do indeed prove all required theorems.

In the course of our formalization, we come to the realization that architects and formal methods experts have different needs from formal models -- not just for \tako{}, but in general. While formal methods experts are concerned with verifiability, architects desire the flexibility to change design features that may improve performance or energy efficiency.

Our work serves the needs of \emph{both} camps.
For \tako{}, our formalisms must account for prefetching and cache replacement policies because they can affect \tako{} correctness.
However, the best prefetching and replacement policies for a desired level of performance and energy efficiency may not be known until late-stage implementation.
Thus, we parameterize our operational model across prefetching policies, cache replacement policies, and network-on-chip specifics so that architects can change them in a \tako{} implementation without compromising their conformance with our MCM. On the formal methods side, we formulate our axioms to be prefix-closed~\cite{kokologiannakis_effective_2018,nienhuis_operational_2016}, a property which enables inductive proofs of implementations against these axioms across all programs.

This work makes the following contributions:

\begin{itemize}
\item \textbf{First Cache-Aware ISA-level MCM.}
We develop the first MCM capable of reasoning about the semantics of cache misses, evictions, writebacks, callbacks, and phantom addresses at an ISA level.
\item \textbf{Parameterized Formal Implementation Model of \tako{}.}
We construct a detailed microarchitectural model of \tako{} in  Dafny \cite{clarke_dafny_2010}. This model parameterizes over \tako{}-adjacent properties that can impact performance (cache replacement policy, prefetching policy, network-on-chip specifics), ensuring that proofs about this model are valid for all choices of these parameters.
\item \textbf{Machine-Checked Soundness Proof of our MCM.} We formally prove that for all programs, any execution of our operational model is also allowed by our ISA-level MCM, ensuring that our ISA-level MCM accurately represents \tako{} functionality. To our knowledge, this is the first end-to-end machine checked proof of an operational implementation against an axiomatic ISA-level MCM.
\item \textbf{General Formal Modeling and Verification Insights.} We discover that architects and formal methods experts have different needs from a formal model, and that a model must serve both communities to be truly effective. We also discover that enforcing prefix-closure~\cite{kokologiannakis_effective_2018,nienhuis_operational_2016} for axioms is extremely useful for enabling inductive proofs of microarchitectural implementations against axiomatic ISA-level MCMs.
\end{itemize}
\section{Background}
\label{sec:background}

\begin{figure}[t]
  \centering
  \includegraphics[width=\linewidth]{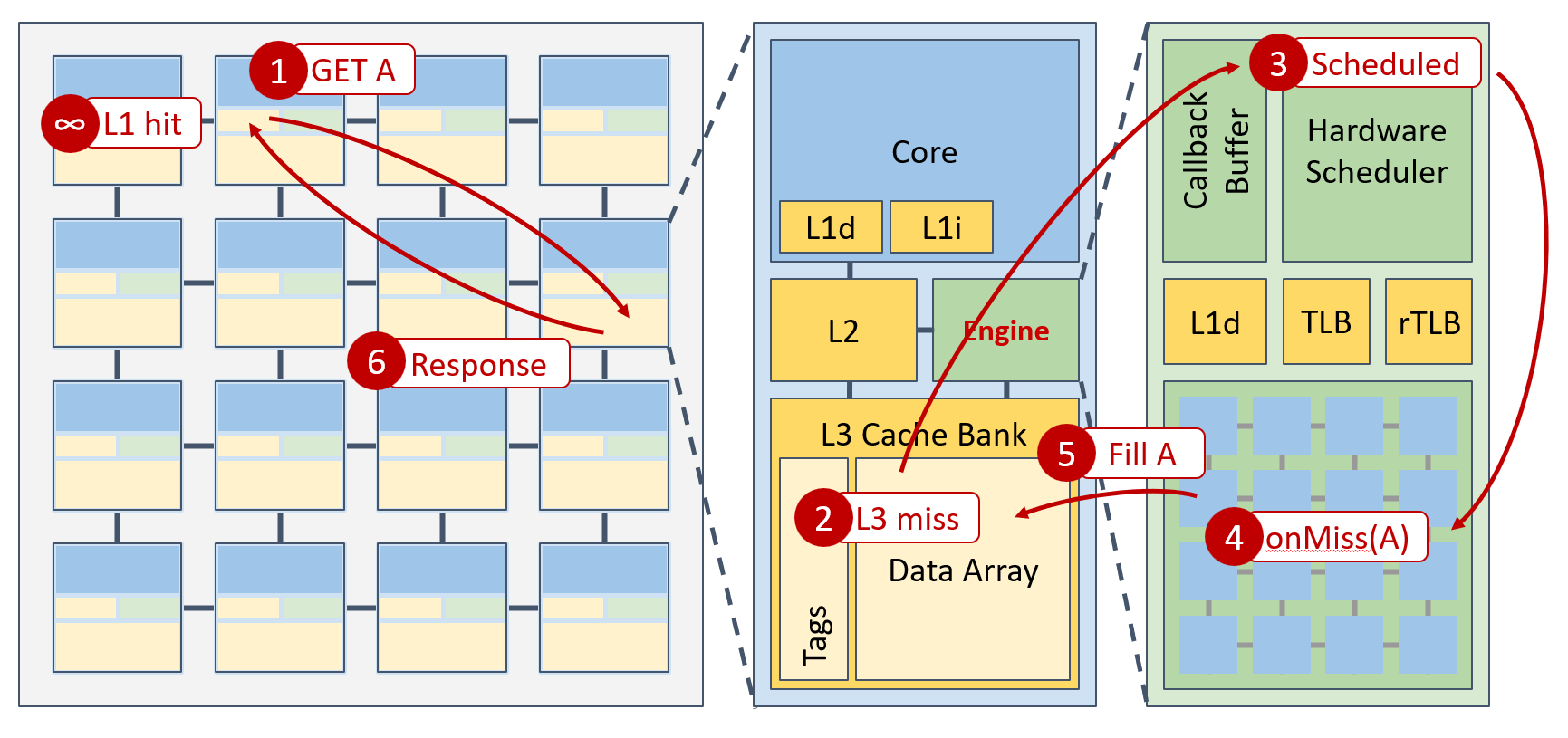}
  \caption{Image from \cite{schwedock_tako_2022} showing the order of events in \tako{} when an \onmiss{} occurs for an L3 phantom address.
  \vspace{-10pt}
  }
  \label{fig:onmiss}
\end{figure}

\subsection{\tako{} Hardware Overview}
As Figure~\ref{fig:onmiss} shows, \tako{} is a tiled chip. Each tile has a core, an L1, a private L2, and a shard of an L3 bank which is shared across tiles. Each tile also has an engine that runs callbacks.

\tako{} allows programmers to register \onmiss{}, \onevict{}, and \texttt{OnWriteback} (henceforth \onwb{}) callbacks for virtual address ranges. These callbacks run on the tile's engine during the corresponding cache events. Figure \ref{fig:onmiss} depicts an \onmiss{} workflow. When an address with a registered \onmiss{} misses in the cache, a hardware thread runs the \onmiss{} on the corresponding engine to calculate the cache line's contents.

Similarly, when a line in the address range is evicted from a cache, an \onevict{} or \onwb{} is invoked to run, depending on whether the data is clean or dirty respectively. Together, \tako{}'s three callback types allow for custom calculations and behavior to run as part of the cache's handling of these address ranges. This functionality allows data transformation to occur as part of data movement instead of occurring once the data has been loaded, and for any clean-up to happen as part of data eviction. Moreover, as the computation results are cached, redundant work is avoided if the same transformation needs to be run again, improving performance on certain workloads. For addresses with no callbacks registered, the semantics of the conventional load/store interface are preserved.

\subsection{Callback Synchronization}
In \tako{}, the microarchitecture's prefetching and cache replacement policies still control \textit{when} a cache line is moved in and out of the cache, as Figure \ref{fig:sample_exec} showed.
Thus, callbacks can be interleaved rather arbitrarily with core thread instructions. While a load reading from an \onmiss{} cannot commit before the \onmiss{} completes, \onevict{} and \onwb{} callbacks can execute anytime after their address is brought into the cache or written to in the cache respectively. \tako{} thus offers a \texttt{FlushRange} synchronization primitive. A \texttt{FlushRange} causes all cache lines with addresses in the mentioned range to be evicted from the cache, invoking their \onevict{} or \onwb{} and blocking the \texttt{FlushRange} till they complete.

\tako{} engines serialize all callbacks to the same address in FIFO order to reduce the possibility of races on those addresses~\cite{schwedock_tako_2022}.
However, \tako{} executions can still easily lead to races and counterintuitive outcomes, as we discuss next.
\section{The Need for Formalization}
\label{sec:formalization}

\begin{figure}[t]
  \centering
  \begin{subfigure}{\linewidth}
  \centering
  \begin{tabular}{|lll|}
    \hline
    \multicolumn{1}{|c|}{Core 0}                       & \multicolumn{1}{c|}{{[}x{]}.\onmiss{}}            & 
    \multicolumn{1}{c|}{{[}x{]}.\onwb{}}       \\ \hline
    \multicolumn{1}{|l|}{\cellcolor[HTML]{C0C0C0}(i1) {[}x{]} $\leftarrow$ 1}  & \multicolumn{1}{l|}{\cellcolor[HTML]{C0C0C0}(i3) {[}x{]} $\leftarrow$ 2} & (i5) {[}y{]} $\leftarrow$ 1 \\
    \multicolumn{1}{|l|}{\cellcolor[HTML]{C0C0C0}(i2) r1 $\leftarrow$ {[}x{]}} & \multicolumn{1}{l|}{\cellcolor[HTML]{C0C0C0}}                            &                             \\
    \multicolumn{1}{|l|}{(i4) r2 $\leftarrow$ {[}y{]}} & \multicolumn{1}{l|}{}                            &                             \\ \hline
    \multicolumn{3}{|c|}{r1=2, r2=0 impossible on \tako{} (if {[}y{]} is initially 0)}                                                                                  \\ \hline
  \end{tabular}
  \vspace{-1pt}
  \caption{}
  \end{subfigure}

  \begin{subfigure}{\linewidth}
      \centering
      \includegraphics[width=\linewidth]{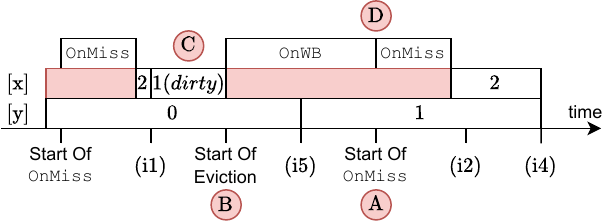}
      \caption{}
  \end{subfigure}
  \caption{(a) An extended version of Figure~\ref{fig:sample_exec}a's program (original program in gray) that depicts the interactions of callbacks and regular address [y]. The outcome r1=2, r2=0 is impossible on \tako{}. (b) A timeline that explains why the outcome is impossible, as r1=2 implies the \onwb{} has completed and written 1 to [y], forbidding r2=0.}
  \vspace{-10pt}
  \label{fig:tako_reasoning}
\end{figure}

\subsection{High Complexity}
\label{subsec:example}

In \tako{}, accesses performed during callbacks can interleave with accesses on core threads, increasing system complexity and counterintuitive outcomes. The confusion is exacerbated by the fact that callbacks can also access regular addresses.
Programmers thus have to reckon not only with phantom address semantics, but also with how cache events for phantom addresses can trigger changes in the values of regular addresses. Next, we provide an example of this complexity.

\subsection{Reasoning About Callbacks}
\label{sec:callback_reasoning}

Consider Figure~\ref{fig:tako_reasoning}a, the example from \S\ref{sec:intro} augmented with an additional \onwb{} callback that writes the value 1 to address [y] (a regular address with no callbacks registered).
If Figure~\ref{fig:tako_reasoning}a were a regular 3-thread program, the outcome r1=2, r2=0 would be possible.
This outcome is actually impossible on \tako{}, but understanding why this is the case requires reasoning about the semantics of caches and callbacks.

Figure \ref{fig:tako_reasoning}b explains this reasoning.
If r1=2, the value for (i2) must have been generated as the result of an \onmiss{} \circled{A} instead of by (i1). Thus, an interspersed \onmiss{} must have run between (i1) and (i2). However, since (i1) would have brought the data for [x] into the cache, that data must also have been evicted \circled{B} before the execution of the \onmiss{} that (i2) read r1=2 from. This eviction must have been an \onwb{}, as the data has been modified by (i1) and is dirty \circled{C}. Due to the serialization of callbacks ensured by the cache controller \circled{D}, we can guarantee that (i5) (the write to [y]) has already completed by the time (i4) runs. Thus, the load of [y] in (i4) must read a value of r2=1 for [y] in this execution, making the outcome r1=2, r2=0 impossible.

\subsection{Formalization To The Rescue}

To enable programmers to rigorously reason about \tako{} programs, we develop an ISA-level MCM for \tako{} (\S\ref{sec:axiomatic}). 
This ISA MCM captures the semantics of cache events and callbacks, but does not require programmers to fully understand \tako{}'s microarchitectural details, thus enhancing \tako{}'s programmability, which we demonstrate with illustrative litmus tests (\S\ref{sec:tako_litmus}).
We ensure that our ISA MCM accurately reflects \tako{} by first constructing a detailed implementation model of \tako{} (\S\ref{sec:operational}), and then proving that all executions of the implementation model are allowed by our ISA MCM (\S\ref{sec:bridge}).
Along the way, we discover insights about how to construct microarchitectural models that are useful for both architects and formal methods experts. We also discover a best practice for ISA-level MCM design to enable effective proofs of hardware implementations against such MCMs.

\section{Cache-Aware Reasoning at the ISA Level}
\label{sec:axiomatic}

\begin{figure}[t]
  \centering
  \begin{subfigure}{\linewidth}
    \centering
    \begin{tabular}{|ll|}
        \hline
        \multicolumn{1}{|c|}{Core 0}                  & \multicolumn{1}{c|}{Core 1} \\ \hline
        \multicolumn{1}{|l|}{(i1) [a] $\leftarrow$ 1} & (i3) r1 $\leftarrow$ [b]     \\
        \multicolumn{1}{|l|}{(i2) [b] $\leftarrow$ 1} & (i4) r2 $\leftarrow$ [a]     \\ \hline
        \multicolumn{2}{|c|}{SC forbids r1=1, r2=0}                                   \\ \hline
    \end{tabular}
    \vspace{-1pt}
    \caption{}
  \end{subfigure}
  \begin{subfigure}{\linewidth}
    \centering
    \includegraphics[scale=0.9]{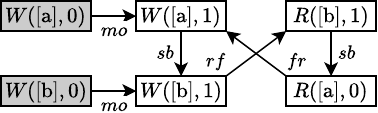}
    \vspace{-5pt}
    \caption{}
  \end{subfigure}
  \caption{(a) The \texttt{mp} (message passing) litmus test. All addresses are assumed to be 0 initially. (b) An execution graph of \texttt{mp} that is outlawed under sequential consistency (SC).}
  \label{fig:axiom_ex}
  \vspace{-15pt}
\end{figure}

\begin{figure*}
\begingroup 
\setlength{\tabcolsep}{4pt}
\renewcommand{\arraystretch}{1.1} 
\begin{tabular}{lr|lr}
$\forall \textbf{R}. \exists ! \textbf{W}. (\textbf{W},\textbf{R}) \in rf$ & \textbf{RfWf1} & $empty([M_s];cbo;cbo;[M_e] \cap thd)$ & \textbf{CboM} \\
$rf \subseteq val \cap addr$ & \textbf{RfWf2} & $empty([E_s];cbo;cbo;[E_e] \cap thd)$ & \textbf{CboE} \\
$\forall A. to(mo, \textbf{W}^A)$ & \textbf{MoWf1} & $empty([\textbf{W}_\textbf{cb}];viscb; [E_s(..,..,false)])$ & \textbf{EvDirty} \\
$mo \subseteq addr$ & \textbf{MoWf2} & $empty(viscb;[E_s(..,..,true)] \setminus [\textbf{W}_\textbf{cb}];viscb)$ & \textbf{WbDirty} \\
$\forall A. to(cbo, CB_{se}^A \cup CB_{me}^A)$ & \textbf{CboWf1} & $empty([M_e];cbo;[M_s] \setminus [M_e];cbo;[E_s];thd;[E_e];cbo;[M_s])$ & \textbf{OEInt} \\
$cbo \subseteq addr$ & \textbf{CboWf2} & $empty([E_e];cbo;[E_s] \setminus [E_e];cbo;[M_s];thd;[M_e];cbo;[E_s])$ & \textbf{OMInt} \\
$viscb \subseteq val$ & \textbf{CboVal} & $\forall M_e. \exists !M_s. (M_s,M_e) \in thd$ & \textbf{OMThd} \\
$[M_s];thd;[M_e] \subseteq cbo$ & \textbf{ThdM} & $\forall E_e. \exists !E_s. (E_s,E_e) \in thd$ & \textbf{OEThd} \\
$[E_s];thd;[E_e] \subseteq cbo$ & \textbf{ThdE} & $empty([M_s];cbo;[M_s] \setminus [M_s];thd;[M_e];cbo;[M_s])$ & \textbf{MeInt} \\
$[E_s];thd;[E_e] \subseteq dirty$ & \textbf{DirtyWf} & $empty([E_s];cbo;[E_s] \setminus [E_s];thd;[E_e];cbo;[E_s])$ & \textbf{EeInt} \\
$irreflexive(hb)$ & \textbf{Hb} & $\forall CB_{me}. \exists !M_e. (M_e,CB_{me}) \in vf$ & \textbf{VfWf} \\
$irreflexive(eco;hb)$ & \textbf{Vis} & $\forall E_s. \exists !M_e. (M_e,E_s) \in ef$ & \textbf{EfWf} \\
\makecell[l]{$irreflexive(rf\, \cup$ \\ \quad ($mo;mo;rf^{-1}) \cup (mo;rf))$} & \textbf{RMW} & \makecell[l]{$\forall Fl. (\forall M_s.(Fl, M_s) \in addr \Rightarrow (Fl, M_s) \in cbo))\, \lor$ \\ \quad $(\exists !E_e. (E_e,Fl) \in eb)$} & \textbf{EbWf} \\
$irreflexive(cbo;hb)$ & \textbf{VisCb} & & \\
\end{tabular}

\vspace{0.5em}
\noindent where \\
\begin{tabular}{lll}
$\textbf{R} = \{R, RMW\}$ & $\textbf{W} = \{W, RMW\}$ & $vf = ([M_e];cbo;[CB_{me}]) \setminus ([M_e];cbo;[CB_{se}];cbo;[CB_{me}])$ \\
$\textbf{R}_{\textbf{cb}} = \{R_{cb}, RMW_{cb}\}$ & $\textbf{W}_\textbf{cb} = \{W_{cb}, RMW_{cb}\}$ & $ef = ([M_e];cbo;[E_s]) \setminus ([M_e];cbo;[CB_{se}];cbo;[E_s])$ \\
$CB_{me} = \{R_{cb},W_{cb},RMW_{cb}\}$ & $CB_{se} = \{M_s, M_e, E_s, E_e\}$ & $eb = ([E_e];cbo;[Fl]) \setminus ([E_e];cbo;[CB_{se}];cbo;[Fl])$ \\
$fr = rf^{-1};mo$ & $eco = (rf \cup mo \cup fr)^+$ & \multicolumn{1}{l}{$sw = ([RMW];rf;[RMW]) \cup ([RMW_{cb}];cbo;[RMW_{cb}])$} \\
\multicolumn{3}{l}{$hb = ((I \times \lnot I) \cup sb \cup sw \cup vf \cup eb \cup ([M_e]; cbo; [E_s]) \cup ([E_e]; cbo; [M_s]))^+$} \\
\multicolumn{3}{l}{$viscb = ([\textbf{W}_\textbf{cb} \cup M_e];cbo;[\textbf{R}_\textbf{cb} \cup E_s]) \setminus ([\textbf{W}_\textbf{cb} \cup M_e];cbo;[CB_{se} \cup \textbf{W}_\textbf{cb}];cbo;[\textbf{R}_\textbf{cb} \cup E_s])$} \\
\multicolumn{3}{l}{$to(R, S) = irreflexive(R) \land transitive(R) \land (\forall s_1, s_2 \in S. R(s_1, s_2) \lor R(s_2, s_1))$} \\
\multicolumn{3}{l}{$race = ((((W \cup R \cup W_{cb} \cup R_{cb}) \times (W \cup R \cup W_{cb} \cup R_{cb})) \setminus ((R \times R) \cup (R_{cb} \times R_{cb}))) \cap addr) \setminus (id \cup hb \cup hb^{-1})$} \\
\multicolumn{3}{l}{$addr$, $val$, $dirty$ denote pairs of events with matching addresses, values, and dirty bits respectively.} \\
\multicolumn{3}{l}{\makecell[l]{Each address is of type \texttt{Synch} or \texttt{Data}. $RMW$/$RMW_{cb}$ run on \texttt{Synch} addresses; $R/W/R_{cb}/W_{cb}$ run on \texttt{Data}.}} \\
\multicolumn{3}{l}{\makecell[l]{$X^A$ is all $X$ events with address $A$. $I$ denotes initialization events. $id$ is the identity relation, i.e. pairs of identical events.}} \\
\end{tabular}
\endgroup
\hrule
\caption{All axioms for our \tako{} MCM. Executions must satisfy all axioms to be allowed. $[A]$ is the elements of type $A$ in relation form~\cite{tarski_calculus_1941}. $A \times B$ are pairs of an element of type $A$ and an element of type $B$. $A \setminus B$ is the elements of $A$ that are not in $B$. Semicolons (;) denote relational composition, e.g., $e1; e2$ is two relations $e1$ and $e2$ where the destination node of $e1$ is the source node of $e2$. $R^{-1}$ is the inverse of $R$. $\exists!$ specifies existence of a \textit{unique} element with the specified property.}
\label{fig:axioms}
\vspace{-20pt}
\end{figure*}

\begin{figure*}
    \centering
    \begin{subfigure}{0.20\linewidth}
        \includegraphics[scale=0.8]{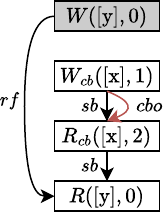}
        \caption{\textbf{Problem:} No callbacks. \\ \textbf{Solution:} \S\ref{sec:phantom_sources}.}
    \end{subfigure}
    \hspace{-0.15cm}
    \begin{subfigure}{0.26\linewidth}
        \includegraphics[scale=0.8]{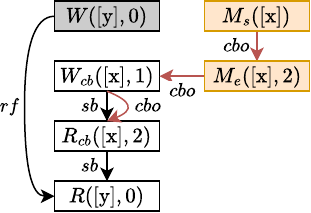}
        \caption{\textbf{Problem:} Value mismatch. \\ \textbf{Solution:} \S\ref{sec:value_correspondence}.}
    \end{subfigure}
    \hspace{-0.7cm}
    \begin{subfigure}{0.27\linewidth}
        \centering
        \includegraphics[scale=0.8]{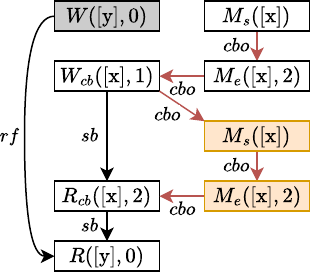}
        \caption{\textbf{Problem:} No intervening eviction. \\ \textbf{Solution:} \S\ref{sec:miss_evict_correspondence}.}
    \end{subfigure}
    \hspace{-0.15cm}
    \begin{subfigure}{0.27\linewidth}
        \centering
        \includegraphics[scale=0.8]{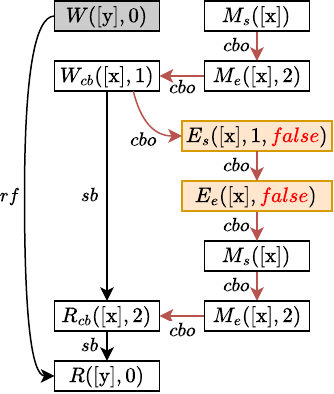}
        \caption{\textbf{Problem:} \onevict{} for dirty data. \\ \textbf{Solution:} \S\ref{sec:miss_evict_correspondence}.}
    \end{subfigure}
    \hfill
    \begin{subfigure}{0.49\linewidth}
        \centering
        \includegraphics[scale=0.8]{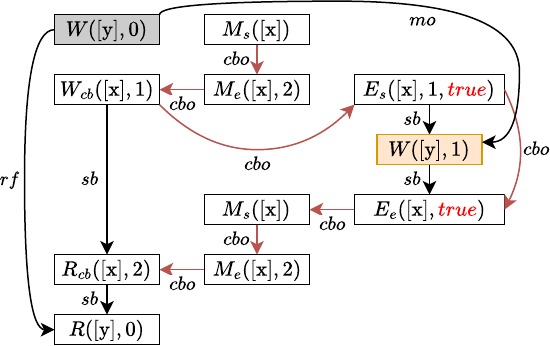}
        \caption{\textbf{Problem:} Violates happens-before visibility for [y]. \\ \textbf{Solution:} \S\ref{sec:happens_before}.}
    \end{subfigure}
    \hfill
    \begin{subfigure}{0.49\linewidth}
        \centering
        \includegraphics[scale=0.8]{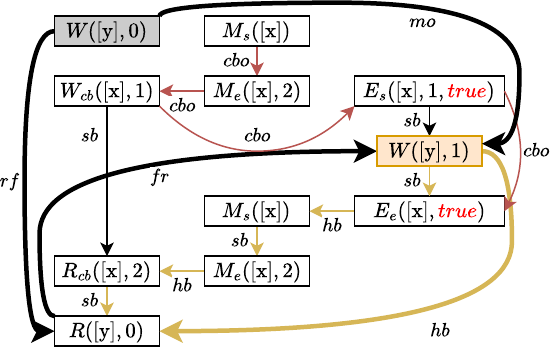}
        \caption{Final execution graph for outcome r1=2, r2=0 \\
        \quad}
    \end{subfigure}
  \caption{(a-e) Faulty candidate execution graphs for Figure \ref{fig:tako_reasoning}a's \tako{} program with an outcome of r1=2, r2=0. (f) The final execution graph for the outcome r1=2, r2=0 once all relevant events, relations, and axioms are added. \S\ref{sec:axiomatic} explains how we encode the semantics of caches and callbacks into axioms to enforce that impossible outcomes like r1=2, r2=0 are forbidden.}
  
  \label{fig:cbo_graph}
  \vspace{-10pt}
\end{figure*}

We now explain our ISA-level MCM for \tako{}.
We begin with background on ISA MCMs (\S\ref{sec:mcm_spec_background}), and then describe our MCM's new events and relations (\S\ref{sec:cache_events}).
The rest of the section discusses key axioms in our MCM in the context of outlawing a forbidden outcome of Figure~\ref{fig:tako_reasoning}a's program.

\subsection{Axiomatic Memory Consistency Models}
\label{sec:mcm_spec_background}

An axiomatic memory model is a formalization used to define the allowable executions of a program. A program execution is represented as a directed graph, where nodes represent instructions and labeled edges encode relations between instructions.
The allowable executions of the program are given by axioms that enforce constraints on these defined relations. Various memory consistency models have been encoded using this approach \cite{alglave_herding_2014, berghofer_better_2009, zhang_ila-mcm_2018, adve_weak_1990, wickerson_automatically_2017, hutchison_fences_2010, lamport_how_1979}.

As an example, consider Figure~\ref{fig:axiom_ex}a's (non-\tako{}) program. Under sequential consistency (SC) \cite{lamport_how_1979}, the allowed executions are those corresponding to an interleaving of each core's instructions in program order. As such, the outcome of r1=1, r2=0 here is outlawed, as r1=1 would indicate (i2) has completed and r2=0 would indicate (i1) has not completed, violating program order on either core 0 or core 1.

Figure~\ref{fig:axiom_ex}b shows an execution graph for the r1=1, r2=0 outcome of Figure~\ref{fig:axiom_ex}a.
All addresses are assumed to initially be 0, as enforced by the initialization writes ($W$) of 0 to [a] and [b].
The other $W$ nodes are the write events ((i1) and (i2)), while the $R$ nodes are the read events. Each node is annotated with its address and value read or written. The $sb$ (sequenced-before) edges connect a given instruction to instructions after it in program order. The $rf$ (reads-from) edge denotes that the read at the target reads from the write at the source. The $fr$ (from-reads) edge denotes that the write at the target occurs \textit{after} the read at the source. 
The $mo$ (modification order) relation establishes a total order on all writes to an address\footnote{In the literature (e.g.,~\cite{alglave_herding_2014}), the relation denoting program order is sometimes labeled $po$ and the relation denoting modification order is sometimes labeled $co$ (coherence order).}.

The axiomatization of SC~\cite{alglave_herding_2014} forbids cycles comprised of the $rf$, $fr$, $sb$ and $mo$ relations. Formally, this is stated as $acyclic(rf \cup fr \cup sb \cup mo)$.
Figure~\ref{fig:axiom_ex}b's execution graph has a cycle comprised of these relations. Thus, it is forbidden under SC, as we would expect for the outcome of r1=1, r2=0.

\subsection{Cache Events And Relations}
\label{sec:cache_events}

Our ISA-level MCM for \tako{} introduces new events and relations to enable reasoning about caches and callbacks. Since the semantics of phantom reads and writes require reasoning about callbacks (e.g., Figure~\ref{fig:sample_exec}), we denote phantom reads and writes using $R_{cb}$ and $W_{cb}$ events
($cb$ for callback). Regular reads and writes are denoted using $R$ and $W$. For both address types, we denote an atomic read-modify-write operation with $RMW_{cb}$ and $RMW$ respectively. We add $Fl$ events to represent the flushing of an address by \texttt{FlushRange}. 

We add events for the beginning and end of each callback, denoted $M_s$ and $M_e$ (\onmiss{} start and end respectively) and $E_s$ and $E_e$ (\onevict{} or \onwb{} start and end respectively). We differentiate between \onevict{} and \onwb{} events using a \emph{dirty bit} for each $E_s$ or $E_e$ event.
The dirty bit is false for \onevict{} events and true for \onwb{} events.

We add a new relation called $cbo$ (callback order) to enforce orderings on these new events. $cbo$ establishes a total order on all callback events ($R_{cb},\, W_{cb},\,RMW_{cb},\,M_s,\, M_e,\, E_s,\, E_e$) for a single address, and reflects \tako{}'s serialization of all callbacks to a given address~\cite{schwedock_tako_2022}.

Next, we describe how the axioms we develop on these relations forbid the r1=2, r2=0 outcome for Figure~\ref{fig:tako_reasoning}a's program. 
Figure~\ref{fig:axioms} contains all our axioms and their names. We refer to axioms using these names throughout the rest of the paper.

\subsection{Ensuring Phantom Address Sources}
\label{sec:phantom_sources}

Figure~\ref{fig:cbo_graph} depicts execution graphs for the r1=2, r2=0 outcome of Figure~\ref{fig:tako_reasoning}a's program. This outcome is impossible on \tako{} due to cache and callback semantics (\S\ref{sec:callback_reasoning}). Figure~\ref{fig:cbo_graph}a depicts an execution graph for this outcome without \tako{}-specific axioms. Traditional MCM axioms cannot reason about cache events and callbacks, so \tako{}-specific axioms are needed to forbid this execution. Graphs (b)-(e) depict execution graphs after the addition of one or more \tako{}-specific axioms and/or relations that outlaw the previous flawed execution graph but not the forbidden outcome of r1=2, r2=0. Graph (f) depicts the execution graph after enough axioms and relations have been added to forbid the r1=2, r2=0 outcome.

First, consider Figure \ref{fig:cbo_graph}a's execution, in which no callbacks execute. The regular read of [y] can read its initial value of 0 from memory, but [x] is a phantom address that does not live outside the cache. Since the caches are initially empty, for [x] to be read, a value for it must first be created in the cache by running an \onmiss{} callback for [x]. Thus, we must outlaw executions in which reads or writes of phantom addresses like [x] are not preceded by the execution of an \onmiss{} for them.

To this end, we add axioms (\textbf{VfWf} in Figure~\ref{fig:axioms}) to ensure that any $R_{cb}$ or $W_{cb}$ must be preceded in $cbo$ by an $M_e$. Analogously, a $Fl$ must be preceded by an $E_e$, denoting that an eviction has completed for the address being flushed (\textbf{EbWf}).
An $Fl$ may also occur before its address is ever brought into the cache (i.e., before there is anything to flush).
These axioms are sufficient to outlaw Figure~\ref{fig:cbo_graph}a's execution.
Adding the required \onmiss{} gives us Figure~\ref{fig:cbo_graph}b, which we discuss next.

\subsection{Ensuring Callback Value Correspondence}
\label{sec:value_correspondence}

Figure~\ref{fig:cbo_graph}b includes an \onmiss{} to generate a value for [x].
However, note that the \onmiss{} generates a value of 2 for [x] (the $M_e([x],2)$ node). This is then overwritten by the write of 1 to [x] (the $W_{cb}$ node). The subsequent read of [x] (i.e., the $R_{cb}$ node) runs after the write to [x], and so should see the updated value of 1. However, it currently reads a 2. 

To fix this problem, we need to add an axiom to ensure that reads of phantom addresses (and eviction callbacks) do not read values that have been overwritten.
Specifically, we must ensure that for a given address $a$, if there is no intervening write (e.g., a $W_{cb}$) to $a$ in $cbo$ order between the end of an \onmiss{} (i.e., an $M_e$) for $a$ and a read (e.g., an $R_{cb}$) of $a$, then the value of the read must match the value of the $M_e$.
On the other hand, if there is an intervening write to $a$ in $cbo$ order between the $M_e$ and the $R_{cb}$ to $a$, then the value of the $R_{cb}$ must match the value of the most recent such write in $cbo$. The latter case occurs when a phantom address is brought into the cache and then written to. Both cases are depicted below:

\begin{center}
 \vspace{-10pt}
 \includegraphics[scale=0.9]{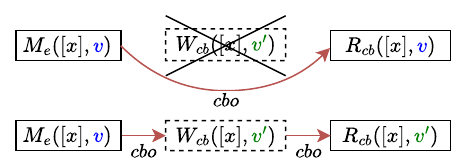}
 \vspace{-7pt}
\end{center}

To express this constraint formally, we first define a $viscb$ relation to denote phantom writes and \onmiss{} results that are visible to phantom reads and eviction callbacks.
We then require that $viscb \subseteq val$ (\textbf{CboVal} in Figure~\ref{fig:axioms}). The $val$ relation links elements with the same value, so this constraint requires that elements linked by $viscb$ must have the same value. This constraint ensures that phantom reads and eviction callbacks must have values that are actually visible to them.

Returning to Figure~\ref{fig:cbo_graph}b, now the only way the read of [x] could get a value of 2 in our program is if another \onmiss{} ran and generated that value of 2, which was then read by the $R_{cb}$ node. Adding this extra \onmiss{} to the graph gives us Figure~\ref{fig:cbo_graph}c, which we discuss in the next section.

\subsection{Ensuring Correct Callback Correspondences}
\label{sec:miss_evict_correspondence}

As Figure~\ref{fig:cbo_graph}c shows, we have now ensured correct values for the phantom reads in our execution. However, note that Figure~\ref{fig:cbo_graph}c contains two \onmiss{} callbacks for [x] without [x] being evicted from the cache in between. This is impossible, and so an eviction callback (\onevict{} or \onwb{}) for [x] must run in between the two \onmiss{} callbacks. 

To enforce this constraint, we require that there must be an \onevict{} or \onwb{} (i.e. an $E_s/E_e$ pair) between the end of an \onmiss{} and the beginning of another \onmiss{} to that same address. Graphically, this constraint requires the dotted events in the diagram below to exist between the $M_s$ and $M_e$ ($thd$ refers to events on the same thread):

\begin{center}
\vspace{-3pt}
 \includegraphics[scale=0.88]{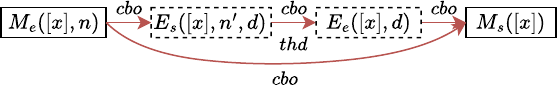}
 \vspace{-6pt}
\end{center}

Formally, this axiom is \textbf{OEInt} in Figure~\ref{fig:axioms}.
We also add a similar axiom for the existence of an \onmiss{} between two eviction callbacks of the same phantom address (\textbf{OMInt}).

The \textbf{OEInt} axiom only ensures that an eviction callback runs in between two \onmiss{}es. It does not enforce whether said callback is an \onevict{} or \onwb{} for the appropriate cases.
With \textbf{OEInt}, an execution graph like Figure~\ref{fig:cbo_graph}d is possible.
Here, we have an \onevict{} for [x] in between the two \onmiss{}es (the dirty bits of the $E_s$ and $E_e$ are false).
Since [x] is written to by the $W_{cb}$ node in the graph before its eviction, an \onwb{} for [x] should run instead of the \onevict{}.
The \onevict{} should only have run if [x] had not been written to in the cache before its eviction.

To this end, we establish a correspondence between the dirty bits of $E_s$ and $E_e$ events and the existence of a write to a phantom address after it is brought into the cache.
If the dirty bit of an $E_s$ is false 
(\onevict{} case), we outlaw the existence of a callback write event in $cbo$ order occurring between the previous $M_e$ (i.e., the end of the most recent \onmiss{}) and this $E_s$. Conversely, if the dirty bit of the $E_s$ is true (\onwb{} case), we necessitate the existence of a write event in between the previous $M_e$ and the $E_s$. Both cases are depicted below:

\begin{center}
 \vspace{-6pt}
 \includegraphics[scale=0.9]{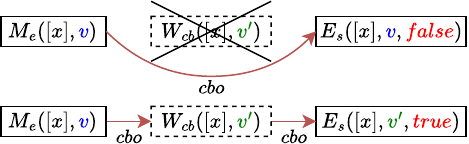}
 \vspace{-7pt}
\end{center}

To express this formally (Figure~\ref{fig:axioms}), we enforce \textbf{EvDirty}
for the \onevict{} and \textbf{WbDirty}
for the \onwb{} cases.

Once we enforce the correct type for each eviction callback, Figure~\ref{fig:cbo_graph}d can no longer be generated. Figure~\ref{fig:cbo_graph}e shows the execution of our running example with an \onwb{} between the two \onmiss{} callbacks for [x], as required. 

While we have now enforced correct values for the callbacks and phantom addresses, the effects of these callbacks and phantom addresses on \emph{regular} addresses have not been enforced. Specifically, Figure~\ref{fig:cbo_graph}e shows that the write of 1 to [y] in the \onwb{} of [x] runs before the $R_{cb}$ of [x], which in turn runs before the read of [y]. (We assume no load-load reordering in our system; the \tako{} paper~\cite{schwedock_tako_2022} uses the x86 ISA which forbids load-load reordering~\cite{berghofer_better_2009}.)
Thus, the read of [y] should see the write of 1 to [y], but there is currently no axiom enforcing this.
To enforce this ordering, we need to augment the happens-before reasoning from traditional MCMs with callback-related orderings, which we do next.

\subsection{Augmenting Happens-Before With Callback Ordering}
\label{sec:happens_before}

\begin{figure}[t]
\centering
\includegraphics[scale=0.9]{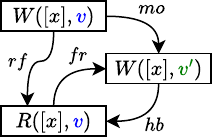}
\vspace{-4pt}
    \caption{A pattern forbidden by traditional happens-before ($hb$).}
    \label{fig:rfmohb}
    \vspace{-12pt}
\end{figure}

\subsubsection{Traditional Happens-Before Ordering}
\label{sec:trad_hb}

In conventional consistency models, the $hb$ (happens-before) relation governs which writes are visible to a read.
The $hb$ relation enables capturing orderings enforced by inter-thread synchronization such as release-acquire~\cite{gharachorloo_memory_1990}.
Stated formally, the $hb$ relation is typically defined as $(sb \cup sw)^+$, or the transitive closure ($+$) of $sb$ and $sw$ (synchronizes with) edges~\cite{adve_weak_1990,batty_mathematizing_2011}. $hb$ also connects initialization events to non-initialization events $(I \times \lnot I)$.

Consider Figure~\ref{fig:rfmohb}'s relational pattern. Here, the read $R$ of [x] reads from the write of $v$ to [x], but a later write of $v'$ to [x] happens-before the read. The read should not observe the older write of $v$ and should instead read the write of $v'$. Thus, this pattern and other similar ones should be forbidden. We, like prior work, do so using Figure~\ref{fig:axioms}'s \textbf{Vis} axiom.

\subsubsection{Adding Callback Ordering to hb}
\label{sec:callback_hb}

Now consider Figure~\ref{fig:cbo_graph}e's execution graph. 
In \tako{}, the write of [y] happens before the read of [y], as \S\ref{sec:miss_evict_correspondence} covers. However, there is no $hb$ edge from the write of [y] to the read of [y] in Figure~\ref{fig:cbo_graph}e, because the traditional $hb$ relation does not take callback orderings into account. Thus, Figure~\ref{fig:rfmohb}'s pattern does not show up in the graph, and the execution is not forbidden by the \textbf{Vis} axiom.
For our $hb$ relation to be accurate, we need to augment it with callback-related orderings.

There are four additions that we make to the traditional $hb$ relation of $(sb \cup sw)^+$ in our \tako{} MCM (Figure~\ref{fig:axioms}). Our first two additions to $hb$ are $vf$ (value-from) and $eb$ (evicts-before) edges. $vf$ edges connect $M_e$ events to the $R_{cb}$ or $W_{cb}$ that they populate the cache for, while $eb$ connects $E_e$ events to the next $Fl$ for the corresponding address. Intuitively, these orderings are part of $hb$ because an \onmiss{} for a phantom address must finish before a read or write to that phantom address, and an eviction of a phantom address must complete before the next \texttt{FlushRange} for that address completes.

The remaining two additions to $hb$ come from \tako{}'s serialization of callbacks to the same address. Specifically, we add any $cbo$ edges from $M_e$ to $E_s$ nodes (i.e., from the end of an \onmiss{} to the start of an eviction for the same address) and from $E_e$ to $M_s$ nodes (i.e., from the end of an eviction to the start of an \onmiss{} for the same address).

Figure~\ref{fig:cbo_graph}f shows the execution graph from Figure~\ref{fig:cbo_graph}e with edges added from our updated $hb$ relation. Our updated $hb$ relation correctly enforces the $hb$ edge we know should exist from the write of [y] to the read of [y].
The graph now contains an instance of Figure~\ref{fig:rfmohb}'s forbidden pattern, which is forbidden by the \textbf{Vis} axiom (\S\ref{sec:trad_hb}). Thus, Figure~\ref{fig:cbo_graph}f's execution is now forbidden by our axioms. There is no additional execution that allows the outcome of r1=2, r2=0, so the axioms we have discussed suffice to forbid that outcome.

\subsection{Summary}

In this section, we showed how we encoded reasoning about \tako{}'s callbacks and cache events into our ISA-level MCM for \tako{}. We could not discuss all our MCM axioms from Figure~\ref{fig:axioms} due to space constraints, but we discussed many important ones.
We have encoded all of our axioms as well as our \tako{} litmus tests in Alloy \cite{jackson_alloy_2002} to support use of our MCM.
A programmer can now simply use our MCM to check whether a given outcome is possible for their \tako{} program, without needing to understand the intricacies of a \tako{} implementation.
\S\ref{sec:tako_litmus} shows how to use our MCM to analyze \tako{} programs.

\section{Programming \tako{} Using Our MCM}
\label{sec:tako_litmus}

We now show how programmers can use litmus tests to write and analyze \tako{} programs using our MCM (\S\ref{sec:no_callbacks}-\conf{\ref{subsec:flush_race}}\ext{\ref{subsec:flushrange_multicore}}).
We also analyze a real \tako{} application (\S\ref{sec:hats}) using our MCM, and highlight the insights we discover about writing \tako{} programs that are both correct and performant.

\subsection{Programs Without Callbacks}
\label{sec:no_callbacks}

Our MCM imposes minimal restrictions on programs without callbacks to avoid adding ordering requirements to non-\tako{} programs.
We do not require preservation of program order between different addresses, so for instance, the execution of \texttt{mp} in Figure~\ref{fig:axiom_ex}b is allowed under our model.

\ext{
\begin{figure}[t]
  \centering
   \begin{subfigure}{\linewidth}
  \centering
  \begin{tabular}{|lll|}
    \hline
    \multicolumn{1}{|c}{Core 0}                       &
    \multicolumn{1}{|c}{Core 1}  &
    \multicolumn{1}{|c|}{\textbf{{[}b{]}.\onmiss{}}} \\
    \hline
    \multicolumn{1}{|l}{(i1) {[}a{]} $\leftarrow$ 1} & 
    \multicolumn{1}{|l}{(i3) \texttt{RMW}({[}b{]}, r1, 1)} &
    \multicolumn{1}{|l|}{\textbf{(i5) {[}b{]} $\leftarrow$ 0}}
    \\
    \multicolumn{1}{|l}{(i2) \texttt{RMW}({[}b{]}, \_, 1)}  & 
    \multicolumn{1}{|l}{(i4) {r2 $\leftarrow$ [}a{]}}  & 
    \multicolumn{1}{|l|}{} \\
    \hline
    \multicolumn{3}{|c|}{\makecell[c]{\texttt{mprmw} ([b] is a regular address, no \onmiss{}): \\ 
    r1 = 1, r2 = 0 forbidden by our MCM}} \\ \hline
    \multicolumn{3}{|c|}{\makecell[c]{\texttt{mpcb} ([b] is a phantom address, \onmiss{} included): \\ 
    r1 = 1, r2 = 0 forbidden by our MCM}} \\ \hline
  \end{tabular}
  \caption{}
  \end{subfigure}
  \begin{subfigure}{\linewidth}
  \centering
  \includegraphics[scale=0.9]{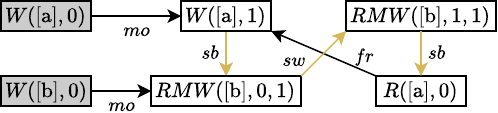}
  \caption{}
  \end{subfigure}
  \begin{subfigure}{\linewidth}
  \centering
  \includegraphics[scale=0.9]{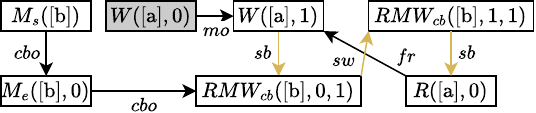}
  \caption{}
  \end{subfigure}
  \caption{(a) The \texttt{mprmw} (no callback) and \texttt{mpcb} (\onmiss{} for [b]) litmus tests. (b) A forbidden execution of \texttt{mprmw} where the $RMW$s augment the $hb$ relation to outlaw the reading of 0 for [a]. (c) An analogous forbidden execution of \texttt{mpcb} showing the same pattern when [b] is a phantom address.}
  \label{fig:mp_extended}
\end{figure}
}

Similar to release consistency~\cite{gharachorloo_memory_1990} and C11~\cite{batty_overhauling_2016}, we classify addresses as data or synchronization addresses, and enforce that if one thread performs a synchronization write that is read by a synchronization read on another thread, accesses after the synchronization read are required to observe accesses before the synchronization write.
In our model's parlance, this synchronization read-write pair induces an $sw$ edge between the two accesses.
We chose RMW operations to implement all accesses to synchronization addresses~\cite{adve:tutorial} for simplicity. (Additional constructs like C11 low-level atomics~\cite{batty_mathematizing_2011} could be added in the future.)
For instance, \conf{if we made [b] a synchronization address and changed}\ext{consider the \texttt{mprmw} litmus test in Figure~\ref{fig:mp_extended}a (where [b] is a regular address and its \onmiss{} is thus omitted). This test changes} the read and write of [b] in Figure~\ref{fig:axiom_ex}b to RMW operations\conf{\footnote{\conf{A read can be replaced by an RMW that writes back the value it read, and a write can be replaced by an RMW that ignores the value it read~\cite{adve:tutorial}.}}}\conf{, they would induce}\ext{, inducing} an $sw$ edge from the write of [b] to the read of [b]\ext{, as shown in Figure~\ref{fig:mp_extended}b}. The \textbf{Vis} axiom then enforces that the read of [a] is required to see the write of 1 to [a], forbidding the outcome of r1=1,r2=0. \textbf{Vis} also enforces per-address SC for regular addresses.

Both regular and phantom conflicting accesses (i.e., a pair of accesses to the same address in different threads not ordered by $hb$ where at least one is a write) constitute a race if they are to a data address, e.g., Figure~\ref{fig:axiom_ex}b's read of [a] and write of 1 to [a] constitute a race. Figure~\ref{fig:axioms} defines our $race$ relation.

\ext{
\subsection{Synchronizing With RMWs To Phantom Addresses}
\label{subsec:rmwcb}
RMWs to phantom addresses (i.e., $RMW_{cb}$ events (\S\ref{sec:cache_events})) can also be used to enforce ordering constraints, similar to how regular address RMWs can be used on conventional programs (\S\ref{sec:no_callbacks}). This is illustrated in the \texttt{mpcb} litmus test (Figure~\ref{fig:mp_extended}a), where the address [b] is a phantom address with an \onmiss{} that returns 0. Similar to \texttt{mprmw}, the outcome of r1=1,r2=0 is forbidden by our \tako{} MCM.

Figure~\ref{fig:mp_extended}c demonstrates a forbidden execution graph that shows the happens-before reasoning in \texttt{mpcb}, which is analogous to that presented in \S\ref{sec:no_callbacks} for \texttt{mprmw}. In this case, as the $cbo$ relation between the two $RMW_{cb}$ events is also added to the $sw$ relation, the same $hb$ edge is constructed between the write and read of [a]. The \textbf{Vis} axiom then forbids the outcome of r1=1,r2=0 for this test, very similar to how it forbids this outcome in \texttt{mprmw}.

Of course, with the use of a phantom address, one must reason about intervening evictions that could occur between the $RMW_{cb}$ accesses. In \texttt{mpcb}, the \textbf{VisCb} axiom ensures that if (i3) reads a value of 1, it must get this value from (i2) (as the \onmiss{} can only produce the value 0), ensuring that no eviction occurs in between. 
}
\ext{
\subsection{Instruction Ordering Within Callbacks}
\begin{figure}[t]
  \centering
   \begin{subfigure}{\linewidth}
  \centering
  \begin{tabular}{|lll|}
    \hline
    \multicolumn{1}{|c}{Core 0} &
    \multicolumn{1}{|c}{{[}x{]}.\onmiss{}}  &
    \multicolumn{1}{|c|}{{[}x{]}.\onwb{}} \\
    \hline
    \multicolumn{1}{|l}{(i1) {[}x{]} $\leftarrow$ 1} & 
    \multicolumn{1}{|l}{(i3) {[}x{]} $\leftarrow$ 0} &
    \multicolumn{1}{|l|}{(i5) {[}a{]} $\leftarrow$ 1}
    \\
    \multicolumn{1}{|l}{} &  
    \multicolumn{1}{|l}{} &
    \multicolumn{1}{|l|}{(i6) {r1 $\leftarrow$ [}b{]}}
    \\ \hline
    \multicolumn{1}{|c}{Core 1}                       &
    \multicolumn{1}{|c}{{[}y{]}.\onmiss{}}  &
    \multicolumn{1}{|c|}{{[}y{]}.\onwb{}} \\
    \hline
    \multicolumn{1}{|l}{(i2) {[}y{]} $\leftarrow$ 1} & 
    \multicolumn{1}{|l}{(i4) {[}y{]} $\leftarrow$ 0} &
    \multicolumn{1}{|l|}{(i7) {[}b{]} $\leftarrow$ 1}
    \\
    \multicolumn{1}{|l}{} &  
    \multicolumn{1}{|l}{} &
    \multicolumn{1}{|l|}{(i8) {r2 $\leftarrow$ [}a{]}}
    \\ \hline
    \multicolumn{3}{|c|}{\makecell[c]{\texttt{icbsb}: r1 = 0, r2 = 0 allowed by our MCM}} \\ \hline
  \end{tabular}
  \caption{}
  \end{subfigure}
    \begin{subfigure}{\linewidth}
    \centering
    \includegraphics[scale=0.8]{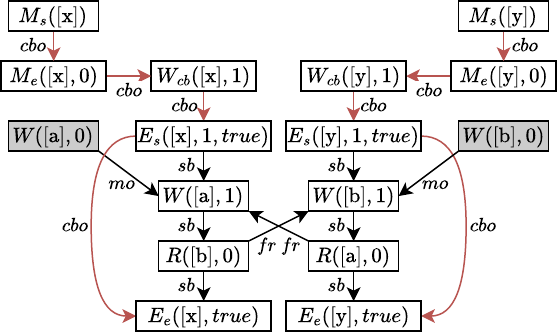}
    \caption{}
    \end{subfigure}
    \caption{(a) The \texttt{icbsb} litmus test. (b) An execution of \texttt{icbsb} that demonstrates the intra-callback instruction reordering that is allowed by our MCM.}
    \label{fig:icbsb}
\end{figure}

Our MCM does not require the preservation of program order between instructions to different addresses in callbacks, similar to how we do not require such ordering in non-\tako{} programs (\S\ref{sec:no_callbacks}).
We demonstrate this using the \texttt{icbsb} litmus test in Figure~\ref{fig:icbsb}a.
In this test, two cores perform writes (i1) and (i2) to phantom addresses [x] and [y] respectively. As there are no restrictions in \tako{} that prevent running callbacks for \textit{different} addresses concurrently, the two \onwb{}s can run at any time with respect to each other, and perform writes and reads to addresses [a] and [b].

The \onwb{}s for [x] and [y] recreate the well-known \texttt{sb} (store buffering) litmus test~\cite{sarkar_understanding_2011}.
As in the \texttt{sb} test, for both loads in the callbacks to return 0 (i.e., the outcome r1=0,
r2=0), the instructions in at least one callback must be reordered.
Figure~\ref{fig:icbsb}b shows an execution with the outcome r1=0,r2=0 that is allowed under our model, thus showing that our model does not require the preservation of program order in callbacks.
More specifically, the axioms in our MCM allow the $fr \cup sb$ cycle among the callback instructions in Figure~\ref{fig:icbsb}b.

By not requiring program order to be preserved in callbacks, our MCM gives computer architects significant freedom when designing the callback engine. In particular, \tako{} designs that buffer and reorder memory operations in the engine can still use our MCM.
}

\subsection{FlushRange as a Synchronization Primitive}
\label{subsec:flush_race}
\begin{figure}[t]
  \centering
   \begin{subfigure}{\linewidth}
  \centering
  \begin{tabular}{|lll|}
    \hline
    \multicolumn{1}{|c|}{Core 0}                       & \multicolumn{1}{c|}{{[}x{]}.\onmiss{}}            & 
    \multicolumn{1}{c|}{{[}x{]}.\onwb{}}       \\ \hline
    \multicolumn{1}{|l|}{(i1) {[}x{]} $\leftarrow$ 1}  & \multicolumn{1}{l|}{(i4) {[}x{]} $\leftarrow$ 0} & (i5) {[}y{]} $\leftarrow$ 1 \\
    \multicolumn{1}{|l|}{\textbf{(i2) {FlushRange[}x{]}}}  & \multicolumn{1}{l|}{} &  \\
    \multicolumn{1}{|l|}{(i3) r1 $\leftarrow$ {[}y{]}} & \multicolumn{1}{l|}{}                            &                             \\ \hline
    \multicolumn{3}{|c|}{\texttt{wbr} (without (i2)): program racy under our \tako{} MCM} \\
    \multicolumn{3}{|c|}{\texttt{wbf} (with (i2)): no race, r1 = 0 forbidden by our MCM} \\ \hline
  \end{tabular}
  \caption{}
  \end{subfigure}
  \begin{subfigure}{\linewidth}
    \centering
    \includegraphics[scale=0.9]{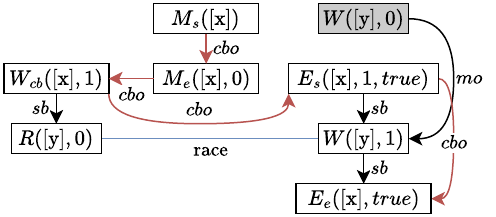}
    \caption{}
  \end{subfigure}
    \begin{subfigure}{\linewidth}
    \centering
    \includegraphics[scale=0.9]{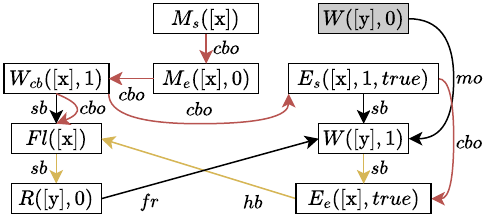}
    \caption{}
  \end{subfigure}
  \caption{(a) The \texttt{wbr} (Writeback Race) and \texttt{wbf} (Writeback Flush) litmus tests. (b) \texttt{wbr} execution where the accesses to [y] race because the \onwb{} is non-blocking. (c) \texttt{wbf} eliminates the race using a \texttt{FlushRange}, and forbids r1=0.}
  \label{fig:wbrace}
  \vspace{-15pt}
\end{figure}

While the use of an \onmiss{}-generated value ($M_e$) by a $R_{cb}$ or $W_{cb}$ induces an $hb$ edge between the events (\S\ref{sec:callback_hb}),
an \onevict{} or \onwb{} for a phantom address can execute anytime after the address is brought into the cache (for \onevict{}) or after the address is written to in the cache (for \onwb{}). This freedom allows races between accesses in these eviction callbacks and those in core program threads.

Consider the \texttt{wbr} litmus test from Figure~\ref{fig:wbrace}a, where the bolded \texttt{FlushRange} (i2) is omitted. This program distills a use case of phantom memory as a write-combining buffer for scatter-updates~\cite{schwedock_tako_2022}, which are published back to regular memory via \onwb{}s. In this test, Core 0 updates the buffer at phantom address [x] in (i1), and (i5) publishes the update to address [y] in regular memory on a writeback. Core 0 also reads the published update in (i3).

Figure~\ref{fig:wbrace}b shows that since the \onwb{} of [x] can execute anytime after (i1), (i5) and (i3) are unordered by $hb$, causing a race.
If we add the \texttt{FlushRange} (i2) from Figure~\ref{fig:wbrace}a to the \texttt{wbr} test to give us the \texttt{wbf} test, the race is eliminated.
Specifically, this \texttt{FlushRange} must either commit before the \onmiss{} of [x] or after the \onwb{} of [x], as enforced by \textbf{EbWf} (Figure~\ref{fig:axioms}).
Committing the \texttt{FlushRange} before the \onmiss{} causes a forbidden cycle in $cbo$ (\textbf{CboWf1}).
Meanwhile, Figure~\ref{fig:wbrace}c depicts the execution where it commits after the \onwb{} of [x]. Here, an $hb$ edge is added between the $E_e$ and the $Fl$ as per Figure~\ref{fig:axioms}'s definitions of $eb$ and $hb$. Combining this edge transitively with the two yellow $sb$ edges
gives us an $hb$ edge between (i5) and (i3), eliminating the race. We now also have a cycle in $hb$ and $fr$ between (i3) and (i5), which violates the \textbf{Vis} axiom, forbidding this execution's outcome of r1=0. Our MCM thus formalizes how \texttt{FlushRange} synchronization can be used to eliminate races.

The race in \texttt{wbr} illustrates a key difference between \tako{} and prior works like IMO~\cite{horowitz_informing_1996} and EcMon~\cite{nagarajan_ecmon_2009} that allow user-space traps for cache events.
In these works, traps effectively have \textit{function call} semantics: they interrupt a core thread (either immediately after a cache event or at a predetermined execution point), execute a handler, and then return control. Thus, the ability of traps to concurrently execute with core threads is greatly reduced if not eliminated. In contrast, in \tako{}, callbacks have \textit{thread} semantics~\cite{schwedock_tako_2022}: they execute on dedicated engines in parallel with core program threads. 
As a result,
conflicting accesses across \tako{} callbacks and core program threads can be races.
This would not be the case in IMO and EcMon where callbacks are not separate threads.

\ext{
\subsection{\texttt{FlushRange} Utility in Multicore Programs}
\label{subsec:flushrange_multicore}
\begin{figure}[t]
  \centering
    \begin{subfigure}{\linewidth}
    \centering
    \begin{tabularx}{\textwidth}{|
        >{\hsize=1.13\hsize\raggedright\arraybackslash}X|
        >{\hsize=1.13\hsize\raggedright\arraybackslash}X|
        >{\hsize=0.74\hsize\raggedright\arraybackslash}X|
    }
    \hline
    \multicolumn{1}{|c|}{Core 0} & \multicolumn{1}{c|}{Core 1} & \multicolumn{1}{c|}{{[}x{]}.\onmiss{}} \\
    \hline
    (i1) \texttt{RMW}({[}x{]}, \_, 1) & (i4) \texttt{RMW}({[}x{]}, \_, 2) & (i7) {[}x{]} $\leftarrow$ 0 \\
    (i2) FlushRange[x]                & (i5) FlushRange[x]                & \\
    (i3) r1 $\leftarrow$ [y]          & (i6) r2 $\leftarrow$ [z]          & \\ 
    \hline
    \end{tabularx}

    \begin{tabularx}{\textwidth}{|l X | l X|}
    \multicolumn{2}{|c|}{[x].\onwb{} (1)} & \multicolumn{2}{c|}{[x].\onwb{} (2)} \\
    \hline
    (i8)  & r3 $\leftarrow$ [y]        & (i12) & if x = 1: \\
    (i9)  & if r3 = 0:                 & (i13) & \quad [y] $\leftarrow$ 1 \\
    (i10) & \quad [y] $\leftarrow$ 1   &       & else: \\
          & else:                      & (i14) & \quad [z] $\leftarrow$ 2 \\
    (i11) & \quad [z] $\leftarrow$ 2   &       & \\
    \hline
    \multicolumn{4}{|c|}{\makecell[c]{
        \texttt{phiR} (with \onwb{} (1)): \\ 
        program racy under our \tako{} MCM \\ 
        \texttt{phiNR} (with \onwb{} (2)): \\ 
        no race, r1 = 0, r2 = 0 forbidden by our \tako{} MCM
    }} \\ 
    \hline
    \end{tabularx}
  \caption{}
  \end{subfigure}
  \begin{subfigure}{\linewidth}
  \centering
  \includegraphics[scale=0.85]{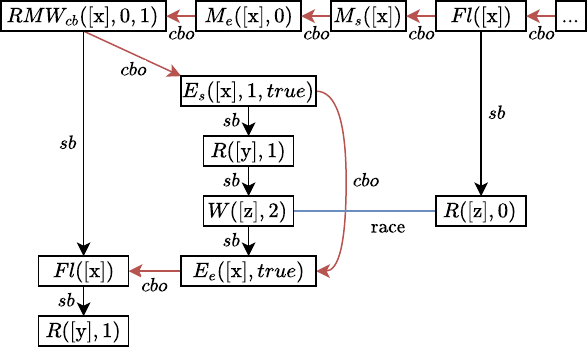}
  \caption{}
  \end{subfigure}
  \begin{subfigure}{\linewidth}
  \centering
  \includegraphics[scale=0.85]{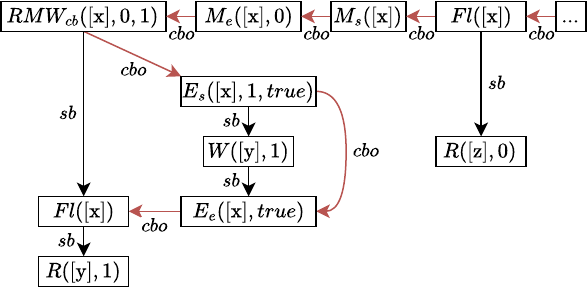}
  \caption{}
  \end{subfigure}
  \caption{(a) The \texttt{phiR} (with \onwb{} (1)) and \texttt{phiNR} (with \onwb{} (2)) litmus tests. (b) Execution snippet showing \texttt{phiR} race due to \onwb{} from core 0's write occurring after \texttt{FlushRange} on core 1. (c) Execution snippet showing how \texttt{phiNR} avoids the race by branching on evicted value in the \onwb{}.}
  \label{fig:phi_extended}
\end{figure}

We now explore the utility of the \texttt{FlushRange} primitive in a multicore setting by considering a multicore version of the \texttt{wbf} litmus test (\S\ref{subsec:flush_race}). 
Figure~\ref{fig:phi_extended} contains 2 implementations of a program (named \texttt{phiR} and \texttt{phiNR}) in which multiple cores concurrently write updates to a phantom address [x] and the \onwb{} of [x] publishes the results back to different locations ([y] and/or [z]), depending on how many updates have previously been published (for \texttt{phiR}) and which update was written last (for \texttt{phiNR}).
This logic is inspired by how the \tako{} paper's acceleration of scatter-updates uses the number of updates in the line at eviction to determine whether to apply the updates in place or log them~\cite{schwedock_tako_2022}.
We additionally use RMW instructions ((i1) and (i4)) to update [x] in both tests because using stores for (i1) and (i4) would result in a race between those two accesses.

In \texttt{phiR}, the program runs with the \onwb{} (1) implementation of the \onwb{} for [x].
In this \onwb{} implementation, (i8) first reads [y]. If [y] has a value of 0 (i.e., if [y] has not yet been written to), (i10) writes 1 to [y]. If [y] has already been written to by a prior \onwb{} for [x], (i11) writes 2 to [z] instead. Thus, the first time the \onwb{} runs, it will write to [y], and if it runs a second time, it will write to [z].

The \texttt{phiR} litmus test is racy, as demonstrated by the execution snippet in Figure~\ref{fig:phi_extended}b. The core cause of the race is that the RMWs (i1) and (i4) can occur in either order, but the \onwb{} will always write to [y] in its first iteration and [z] if it runs a second time.
Consider the case where (i4) runs first (not shown in Figure~\ref{fig:phi_extended}b) and then (i5) invokes the \onwb{} and causes [y] to be updated to 1 (this \onwb{} is also not shown). Figure~\ref{fig:phi_extended}b shows that when (i1) subsequently writes to [x] and (i2) then invokes the \onwb{}, this will trigger the write to [z] in (i11). However, nothing stops this write from racing with the read of [z] in (i6) on core 1, giving us a race.

This execution demonstrates a key requirement when using \texttt{FlushRange} for synchronization: for \texttt{FlushRange} to be able to eliminate a race, callbacks should not be able to run accesses that cause the race after the \texttt{FlushRange} in question has committed.
Here, it is possible for the \onwb{} to be triggered and run (i11) after (i5) has committed, so (i5) is unable to prevent (i11) from racing with (i6).

The above requirement is fulfilled by the \texttt{phiNR} litmus test from Figure~\ref{fig:phi_extended}a that uses \onwb{} (2) as the \onwb{} for [x]. Here, the \onwb{} uses the evicted value to determine which value to write.
(i12) checks the evicted value.
If it is 1 (i.e., the last update was from Core 0), then (i13) writes 1 to [y].
If it is 2 (i.e., the last update was from Core 1), then (i14) writes 2 to [z].
Thus, once (i5) commits, any write to [z] is guaranteed to have completed -- the \onwb{} only writes 2 to [z] if [x] is 2, and that can only happen after (i4) and before (i5) commits.
Thus, there is no write to [z] that can race with (i6) in any execution.
(Similarly, once (i2) commits, there is no write to [y] left that can race with (i3), eliminating races on [y] as well.)
Figure~\ref{fig:phi_extended}c shows how there is no write to [z] in an \onwb{} triggered by (i1) and (i2), thus eliminating the race seen in Figure~\ref{fig:phi_extended}b.

In \texttt{phiNR}, the outcome r1=0,r2=0 is forbidden by our MCM, because neither (i3) nor (i6) can run before at least one \texttt{FlushRange} instruction (i.e., (i2) or (i5)) commits. Since the \texttt{FlushRange} instructions are each after RMW instructions that write to [x] in program order, the first \texttt{FlushRange} to commit will cause the \onwb{} to run, which will update one of [y] or [z]. Thus, at least one of [y] or [z] will have been written to before (i3) or (i6) run, preventing them both from returning 0 and forbidding the outcome of r1=0,r2=0.

In \texttt{phiR} and \texttt{phiNR}, the \onwb{} callback runs a maximum of two times, so the complexity is relatively easy to manage.
Next, we investigate \tako{}'s implementation of accelerated graph traversal, which is more difficult to write correctly because it has writes in \onevict{} callbacks that can run an arbitrary number of times.
}

\subsection{A \tako{} Application Case Study: HATS}
\label{sec:hats}

\begin{figure}[t]
  \centering
  \begin{subfigure}{\linewidth}
  \centering
  \begingroup
  \setlength{\tabcolsep}{4pt}
  \begin{tabular}{|llllll|}
    \hline
    \multicolumn{2}{|c|}{Core 0}                       & \multicolumn{2}{c|}{{[}e{]}.\onmiss{}}            & \multicolumn{2}{c|}{{[}e{]}.\onevict{}}       \\ \hline
    (i1) & \multicolumn{1}{l|}{\texttt{RMW}({[}e{]}, r1, 1)} & 
    (i4) &
    \multicolumn{1}{l|}{r3 $\leftarrow$ [g]} & \textbf{(i9)} & \textbf{if [e] $\neq$ 1:} \\
    (i2) & \multicolumn{1}{l|}{FlushRange{[}e{]}} & (i5) & \multicolumn{1}{l|}{if r3 $\neq$ 1:} & (i10) & \quad {[}$\ell${]} $\leftarrow$ 1 \\
    (i3) & \multicolumn{1}{l|}{r2 $\leftarrow$ {[}$\ell${]}} & (i6) & \multicolumn{1}{l|}{\quad [g] $\leftarrow$ 1} &  &  \\
     & \multicolumn{1}{l|}{} & (i7) & \multicolumn{1}{l|}{\quad [e] $\leftarrow$ 0} & & \\
    & \multicolumn{1}{l|}{} &  & \multicolumn{1}{l|}{else:} & & \\
    & \multicolumn{1}{l|}{} & (i8) & \multicolumn{1}{l|}{\quad [e] $\leftarrow$ 1} & & \\ \hline
    \multicolumn{6}{|c|}{\texttt{hatsR} (without (i9)): program racy under our \tako{} MCM} \\
    \multicolumn{6}{|c|}{\texttt{hatsNR} (with (i9)): no race, r1 $\neq$ r2 forbidden by MCM} \\ \hline
  \end{tabular}
  \endgroup
  \caption{}
  \end{subfigure}
    \begin{subfigure}{\linewidth}
    \centering
    \includegraphics[scale=0.9]{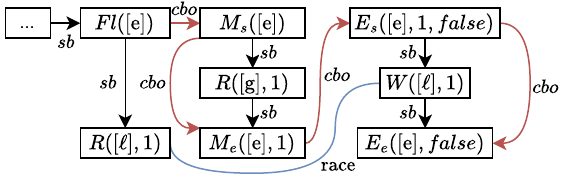}
    \caption{}
  \end{subfigure}
    \begin{subfigure}{\linewidth}
    \centering
    \includegraphics[scale=0.9]{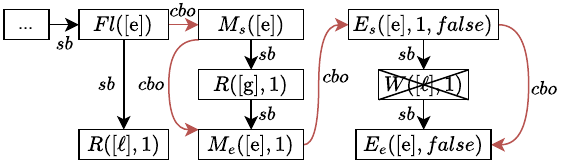}
    \caption{}
  \end{subfigure}
  \caption{(a) \texttt{hatsR} and \texttt{hatsNR} litmus tests. (b) Execution snippet showing \texttt{hatsR} race due to \onevict{} writing to the log post-traversal. (c) Snippet showing how \texttt{hatsNR} eliminates the race by only logging valid edges in the \onevict{}.}
  \label{fig:hats}
  \vspace{-10pt}
\end{figure}

We now analyze \tako{}'s implementation of HATS (Hardware-Accelerated Traversal Scheduling)~\cite{mukkara_exploiting_2018}, which accelerates graph processing.
The \tako{} paper~\cite{schwedock_tako_2022} states that \onmiss{} and \onevict{} should be side-effect-free because they can occur at any time, but does not formalize why.
\tako{}'s HATS implementation ignores this guideline, making it an interesting case study.
We use our MCM to analyze HATS and identify when \onevict{} side effects are acceptable.

\tako{}'s HATS implementation creates a phantom address range to store all graph edges \emph{sequentially}. It then uses \onmiss{} callbacks to traverse the graph and place edges in this range as the main program thread accesses the phantom addresses. Thus, the main thread can iterate over the edges sequentially (an easy-to-predict pattern with
good locality), while the engine handles the hard-to-predict graph traversal.

Each time the \onmiss{} runs, it provides the next edge from the traversal. The \onmiss{} only returns each edge once. Even if the same phantom address misses twice, the two \onmiss{}es will return different edges. Thus, edges that are evicted before being processed would be lost without additional handling. 
Eviction callbacks therefore log edges that are evicted too early to regular memory.
The main thread processes the log after iterating through the phantom addresses~\cite{schwedock_tako_2022}.

The \texttt{hatsR} litmus test (Figure~\ref{fig:hats}a, without (i9)) contains the core logic of a naive \tako{} HATS implementation.
The phantom address [e] represents a graph edge, with 0 being a valid edge and 1 an invalid one. Core 0 reads [e] to process it using an RMW, atomically reading [e] and writing 1 to [e] to mark it as processed. This ensures that an eviction cannot occur between reading the edge and marking it processed, which would cause the edge to be processed twice. Core 0 then flushes [e] to ensure that any in-progress \onevict{}s for it are completed. It then processes the log (represented by reading address [$\ell$]). The \onmiss{} maintains an engine-local view of the graph (address [g]), and populates [e] with a valid edge if traversing the graph (i.e., if [g] = 0) or an invalid edge if the graph has already been traversed (i.e, if [g] = 1). The \onevict{} writes the edge to the log ([$\ell$]). The \onwb{} is empty. The intuition here is that if the edge is evicted before being written to, it has not yet been processed.

This \texttt{hatsR} program is racy, as Figure~\ref{fig:hats}b's graph snippet shows.
The specific problem is that an \onmiss{}-\onevict{} sequence for a phantom address can run as many times as the cache wants, without action from the core. For instance, the cache's prefetcher may load [e], the cache may then evict it, and this sequence may then repeat.
Here, [e] is first loaded in and processed (or not) by the core, and evicted or written back (depending on whether or not it was processed) before
(i2)'s \texttt{FlushRange}.
However, nothing stops the cache from then \emph{reloading} [e] (triggering a second \onmiss{}) and then evicting it (triggering an \onevict{}). This second \onmiss{}-\onevict{} sequence is shown in Figure~\ref{fig:hats}b. The write in the \onevict{} races with the read of the log in (i3). While this write may not change the value of [$\ell$] in the test (if [e] was not processed by the core, [$\ell$] will already be 1), recall that each such write to [$\ell$] corresponds to updating the log in the real HATS application, and would cause spurious edges to be added to the log, impacting correctness.

Fixing this issue is tricky as the cache cannot be prevented from loading or evicting phantom addresses after the graph traversal completes.
The solution is to ensure that once the traversal completes, (i) the \onmiss{} does not provide spurious data for [e] and (ii) neither the \onmiss{} or \onevict{} access addresses besides their registered address [e] and engine-local addresses like [g]. This solution is presented in the \texttt{hatsNR} test (Figure~\ref{fig:hats}a including (i9)), where the \onevict{} only logs valid edges, and the \onmiss{} (as before) returns invalid edges once the graph has been traversed. Figure~\ref{fig:hats}c shows how this eliminates the race: post-traversal \onevict{}s will never log edges in \texttt{hatsNR}, thus eliminating the write of the race. We encoded \texttt{hatsNR} in our Alloy model of our MCM and searched for races with a large bound. None were found, giving us confidence that \texttt{hatsNR} is indeed race-free.

\section{A Parameterized Model of \tako{}}
\label{sec:operational}

\subsection{Operational Models}

An operational model is a transition system,
which consists of a set of states $S$, as well as two predicates over elements of this set: an initial state predicate $Init(s)$ that is true when $s \in S$ represents a valid initial state, and a transition predicate $Next(s, s')$ that is true for a pair of states $s, s' \in S$ when they represent a valid transition from $s$ to $s'$.

We implement our transition system in Dafny \cite{clarke_dafny_2010}, a verification-aware language that enables proving properties about programs and abstract models in a partially automated way. Dafny converts properties into Satisfiability Modulo Theories (SMT) queries and then uses Z3~\cite{de_moura_z3_2008} to verify them.

\subsection{\tako{} State Machine Overview}
\label{subsec:state_machine}

\begin{figure}[t]
    \centering
    \includegraphics{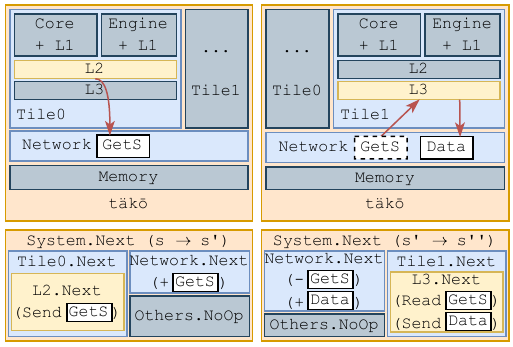}
    \caption{Our hierarchical transition system instantiated with two tiles, executing two transitions ($s \rightarrow s'$ and $s' \rightarrow s''$). In the first transition, \texttt{Tile0}'s \texttt{L2} cache sends a \texttt{GetS} request into the \texttt{Network}. In the second, \texttt{Tile1}'s \texttt{L3} cache receives said request and responds with a \texttt{Data} message. All other state machines perform a \texttt{NoOp} transition (colored \fcolorbox{nopColourborder}{nopColour}{\rule{0pt}{6pt}\rule{6pt}{0pt}}\,).}
  \label{fig:sm_overview}
  \vspace{-10pt}
\end{figure}

Figure~\ref{fig:sm_overview} depicts the components in our \tako{}
transition system, 
instantiated
with two tiles,
as well as two example transitions.
The 
model
is hierarchical: each tile of the \tako{} chip (Figure \ref{fig:onmiss}) is its own transition system, consisting of individual transition systems representing each major component of said tile (the \texttt{Core + L1}, \texttt{Engine + L1}, \texttt{L2}, and \texttt{L3} slice). A single \texttt{Memory} state machine represents main memory.

To enable communication between the individual components, we add a state machine called \texttt{Network} to the design, which contains all in-flight messages between components. During any individual component transition, a single message may be read from or written into the \texttt{Network} (potentially both). The \texttt{Network}'s internal structure determines which messages can be delivered during a transition: for coherence messages, this is an unordered set, allowing arbitrary reordering to overapproximate various network-on-chip designs. For engine requests, the \texttt{Network} is stricter, enforcing FIFO ordering per address on callback requests, as \tako{} requires~\cite{schwedock_tako_2022}.

The overall $Next$ predicate for the system is a transition of a single component state machine alongside a transition of the \texttt{Network} to capture a potential message the component transition might send and/or receive. Figure \ref{fig:sm_overview} illustrates transitions in greater detail, depicting how the transition system represents the behavior of the \texttt{L2} requesting data from the \texttt{L3}. In this case, the \texttt{L2.Next} predicate adds a \texttt{GetS} to the \texttt{Network}, and a subsequent transition in the \texttt{L3} state machine receives this message from the \texttt{Network} and replies with the requested data. Thus, an execution trace for a program in the full state machine is decomposed into individual atomic steps that are taken by its component state machines, alongside \texttt{Network} steps for communication. 

Our model currently supports programs with \texttt{Load}, \texttt{Store}, \texttt{RMW}, \texttt{Flush}, and \texttt{Branch} instructions. The execution of an instruction potentially corresponds to several transitions that occur in the system (e.g., to request its value from the memory hierarchy).
Each instruction is committed using a \texttt{PerformInst} transition in either the \texttt{Core} or \texttt{Engine} component state machines, depending on if the instruction is in a program thread or a callback.

Our caches are inclusive (enforced by \tako{}~\cite{schwedock_tako_2022}), and kept coherent through a hierarchical directory-based MSI protocol~\cite{nagarajan_primer_2020}.
The higher protocol has the \texttt{L2} as a directory and the \texttt{Core} and \texttt{Engine} L1s as children, while the lower protocol has \texttt{L3} shards as a directory and \texttt{L2} caches as children.
We use the HieraGen~\cite{oswald_hieragen_2020} approach of a proxy L2 cache to communicate between the two protocols.
Our protocols do not require point-to-point ordering, as our \texttt{Network} model does not enforce point-to-point ordering for coherence messages.
 
\subsection{Model Parameterization} 
\label{subsec:model_param}
Our model is parameterized over the size of all caches, the executing program, the number of cores, and the mapping of addresses to L3 banks. By making these parameters generic instead of concrete values, we ensure that when we prove facts about the transition system (\S\ref{subsec:induction_primer}), we in fact prove them for \emph{any} configuration of these values. This parameterization makes our proof notably harder, as we now cover a wider range of possibilities. However, our proof becomes much more useful, as it applies to \emph{all} configurations of these parameters.

\subsection{Environmental Transitions}
\label{subsec:env_trans}

\begin{figure}[t]
    \centering
    \includegraphics[width=\linewidth]{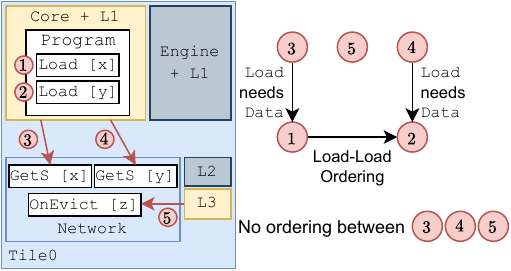}
    \caption{Environmental transitions in action. For the five pictured potential transitions, \circled{3}, \circled{4}, and \circled{5} can occur at any time without any dependencies on each other. The \texttt{Load}s (\circled{1}, \circled{2}) are dependent on the \texttt{Data} being in their cache.}
  \label{fig:env_trans}
  \vspace{-15pt}
\end{figure}

In systems like \tako{}, callback code can execute due to events like evictions or prefetches.
Hardware controls when these events happen, so they may interleave arbitrarily with main program threads.
To model such variable timing, we use environmental transitions (transitions that are not dependent on instructions)~\cite{wickerson_remote-scope_2015} to overapproximate cache behavior.
Environmental transitions decouple memory hierarchy transitions from the instructions on cores and engines, enabling us to model cases where an instruction triggered a memory request, as well as cases where the same request was triggered by a prefetch or eviction.
Using environmental transitions thus ensures that our proof of consistency (\S\ref{sec:bridge}) is sound even under varied prefetching and cache replacement policies.

Figure~\ref{fig:env_trans} illustrates environmental transitions in the context of two loads.
Consider modeling a load's execution if its data is not present in the L1, e.g., the load of [x] in Figure~\ref{fig:env_trans}.
Instead of making the load send a \texttt{GetS} to the \texttt{L2} state machine, we decompose this transition into two independent ones (\circled{1} and \circled{3} in Figure~\ref{fig:env_trans}).
\circled{1} is a \texttt{PerformLoad} step that can only execute if the data is in the L1, and \circled{3} is a \texttt{SendGetS} step that can execute \emph{whenever the data is not in the cache}. We thus capture executions in which a \texttt{GetS} is triggered without a specific load causing it (e.g., on a prefetch), as well as executions where the events are causally linked.
Thus,
even though Figure~\ref{fig:env_trans} has the load of [x] (\circled{1}) before the load of [y] (\circled{2}) in program order, their requests to the memory hierarchy (\circled{3} and \circled{4}) are not ordered with respect to each other due to environmental transitions, as they could be prefetched out of order. Callback scheduling and running are also environmental transitions, since misses, evictions, and writebacks can occur arbitrarily.
Thus, the eviction of [z] (\circled{5}) in Figure~\ref{fig:env_trans} can also be interleaved with other transitions arbitrarily.

We allow callback environmental transitions to repeat if their preconditions are met.
For instance, our model can execute repeated \onmiss{}-\onevict{} sequences for an address [x] without a core ever requesting [x].
This is because a prefetch could bring [x] in and the cache could then evict it at any time. Allowing such loops enables our model to overapproximate replacement policy or prefetcher-based triggering of callbacks that might cause unexpected outcomes.
Our proof in \S\ref{sec:bridge} is valid across all such callback combinations.

\section{A Machine-Checked Consistency Proof}
\label{sec:bridge}

Here, we describe how we produce a machine-checked, all-program proof that our \tako{} ISA-level MCM (\S\ref{sec:axiomatic}) is sound with respect to our model of the \tako{} hardware (\S\ref{sec:operational}). 

\subsection{Proof by Induction}
\label{subsec:induction_primer}
Induction is a powerful proof technique that can prove properties about infinite executions of a state machine while only reasoning about one transition at a time. In the realm of hardware, where many designs are modeled as state machines, this approach is broadly applicable \cite{sheeran_checking_2000, 9716812, wang_specification_2023, peled_large_2025}.

Given a state machine and a property $P(s)$,
induction requires us to prove two things: a) that the initial state satisfies $P(s)$, and b) that all transitions preserve $P(s)$. This proves that $P(s)$ is maintained throughout the execution. 

Formally, the second obligation is $P(s) \wedge Next(s, s') \implies P(s')$), i.e., if we are in a state satisfying $P$, any transition from that state should lead to another state that satisfies $P$. Alas, this proposition simply does not hold for most systems.

\begin{figure}[t]
  \centering
  \vspace{-2pt}
  \includegraphics[scale=0.85]{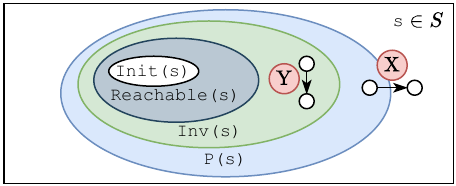}
  \caption{A Venn diagram motivating inductive invariants. Non-reachable states can leave the $P(s)$ set by transitioning (\circled{X}), so a strengthening $Inv(s)$ is constructed such that any transition of a state in $Inv(s)$ (\circled{Y}) remains in $Inv(s)$.}
  \label{fig:ind_inv}
  \vspace{-13pt}
\end{figure}

To understand why, consider Figure~\ref{fig:ind_inv}'s Venn diagram. The outer rectangle represents all possible states in the system, including states not reachable by the state machine. The outer circle represents $P(s)$; i.e., all states $s$ for which $P$ holds. It is possible that a transition like \circled{X} exists, such that the above implication is violated. This \emph{does not} mean that our system is incorrect, as \circled{X} originates in an unreachable state and is thus a spurious counterexample. But it \emph{does} mean we cannot directly use $P$ in our inductive proof.

Instead, we use an \textit{inductive invariant} ($Inv(s)$ in Figure~\ref{fig:ind_inv}), a strengthening of $P(s)$ (i.e., $Inv(s) \implies P(s)$) which \textit{is always preserved} under $Next$ (as exemplified by \circled{Y}). Our proof obligation is then to show that $Inv(s) \wedge Next(s, s') \implies Inv(s')$ and $Init(s) \implies Inv(s)$. This ensures that $P(s)$ holds throughout the execution, because $Inv(s) \implies P(s)$. Finding an inductive invariant for a system is a key challenge when verifying a system inductively~\cite{zhang_basilisk_2025,padon_ivy_2016,ma_i4_2019}.

\subsection{Intermediate State Machine}
\label{subsec:inter_state_machine}

We wish to demonstrate via induction that each axiom in our MCM holds for all executions.
However, recall that axioms are properties of execution graphs, meaning they hold for a \textit{full execution} of a program. In contrast, our operational model is a transition system that \textit{incrementally builds the execution} with each transition. Inductive proofs (\S\ref{subsec:induction_primer}) require a property to hold during these intermittent stages as well.
Thus, we first need a notion of a \textit{partial execution graph} which represents the execution graph of a program that has not yet completed running. 
To that end, we first build an intermediate state machine (henceforth ISM for short) that is much simpler than the operational model in \S\ref{sec:operational}. The state of this \ism includes a partial execution graph for a program, and its transitions represent how running a program updates this graph.
We then require that each axiom, when strengthened with an inductive invariant (\S\ref{subsec:induction_primer}), is true for the partial graph at $Init(s)$, and is preserved by $Next$'s additions to these partial graphs.

\subsection{Using Prefix-Closure To Ensure Provable Axioms}
\label{subsec:prefix_closure}
A key potential pitfall with this inductive approach is that certain axioms can be true about the final, \emph{full} execution graph, but fail to hold for partial graphs along the way. For example, an axiom we originally added to our model expressed the property that for each $M_s$ node in the graph, there was a corresponding $M_e$ that belonged to the same thread.
An axiom like this was needed to outlaw \textit{full} executions where there were more $M_s$ events than $M_e$ events; such executions are impossible, as an \onmiss{} requires both an $M_s$ and an $M_e$.

\begin{figure}[t]
  \centering
  \includegraphics[width=\linewidth]{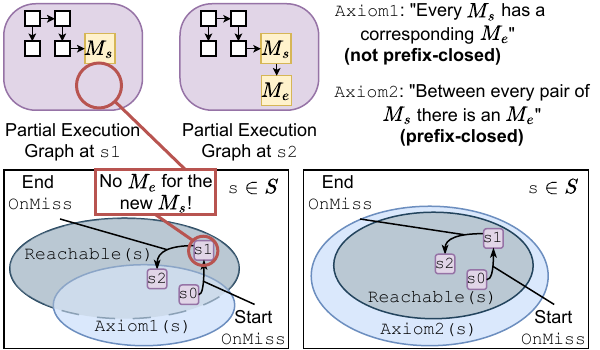}
  \caption{Two formulations of an axiom restricting the number of $M_s$ and $M_e$ events in an execution graph. While both are true about full executions of our operational model, $Axiom1$ is not provable using induction, because there are transitions from reachable states (e.g., $s0 \rightarrow s1$) where $Axiom1$ temporarily fails to hold. $Axiom2$ (\textbf{MeInt} in our MCM), phrased in a prefix-closed manner, avoids this issue by ensuring that the axiom holds for all reachable partial executions as well.}
  \label{fig:non_inductive}
  \vspace{-10pt}
\end{figure}

Figure~\ref{fig:non_inductive} illustrates why this axiom ($Axiom1$) cannot be verified \emph{despite being true}. Consider the transition which adds an $M_s$ to the \ism's partial execution graph, representing an \onmiss{} starting in the engine. As this callback has not completed, its associated $M_e$ has not yet been added. As such, while $Axiom1$ is valid for full executions, it \emph{cannot be proven} by standard induction as it fails to hold for a partial execution graph after a transition from a reachable state ($s0 \rightarrow s1$).

To remedy this, a key feature of the axiom set we design in \S\ref{sec:axiomatic} is that each axiom is prefix-closed: that is, if it holds for the full execution, it will hold for all partial executions along its construction (its \emph{prefixes})\footnote{Prefix-closure in the literature is typically defined with respect to a commitment order: our commitment order respects $cbo$, meaning we never add callback events to the graph in an order that violates $cbo$.}. This property has a remarkable consequence for inductive verification: if an axiom is prefix-closed and true about the system, \textit{one can always find an inductive invariant to prove it}. For our system, we changed $Axiom1$ to $Axiom2$, which states that for every pair of $M_s$, there exists an $M_e$ in between them (\textbf{MeInt} in Figure \ref{fig:axioms}). This accomplishes the same goal, but is also true when an \onmiss{} is partially complete, and thus can be verified inductively.

Prior work~\cite{nienhuis_operational_2016, kokologiannakis_effective_2018} has used prefix-closure for its axiom sets to ensure that its construction of partial execution graphs starts and remains consistent as it produces a complete execution trace. We leverage this same property in a new way: to ensure axioms about our system, when true about full executions, are always provable on a state machine representation of the hardware using inductive verification. This approach can be applied for general proofs of hardware correctness against prefix-closed MCM axioms. As such, future designers of MCMs should strive for axioms that respect this property to unlock this method of MCM verification. 

\subsection{Proof Architecture: Dividing the Proof Obligation}
\label{subsec:proof_arch}

\begin{figure*}
  \centering
  \includegraphics[width=\linewidth]{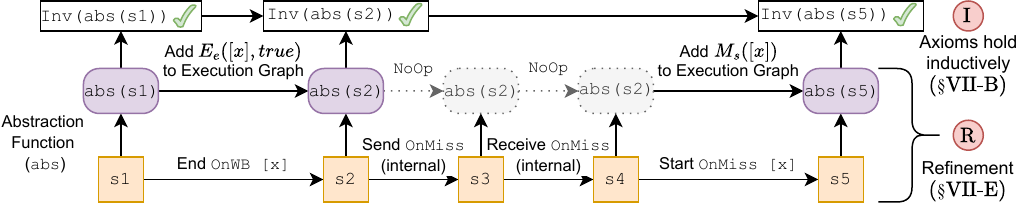}
  \vspace{-15pt}
  \caption{The two components of our soundness proof. \circled{R} The refinement proof (\S\ref{subsec:refinement}) proves that each operational model execution (e.g., $s1 \rightarrow s5$), abstracted at each state, 
  maps to an intermediate state machine execution (e.g., $abs(s1) \rightarrow abs(s5)$). Operational model transitions either add nodes to the partial execution graph (e.g., $s1 \rightarrow s2$) or are internal steps that abstract up to \texttt{NoOp}s (e.g., $s2 \rightarrow s3$). \circled{I} The inductive proof that the axioms hold as the partial execution graph grows (\S\ref{subsec:inter_state_machine}).}
  \label{fig:proof_anatomy}
  \vspace{-10pt}
\end{figure*}

Figure~\ref{fig:proof_anatomy} illustrates the two forms of reasoning in our end-to-end proof. First, one must reason (\circled{I}) about the \ism itself (\S\ref{subsec:inter_state_machine}) to determine if it satisfies the axiom we are verifying using induction (\S\ref{subsec:induction_primer}). Second, one must connect (\circled{R}) the \ism with the operational model to ensure that the \ism is a faithful representation of the hardware.

Previous work has separated these two forms of reasoning by introducing an intermediate state machine that builds partial execution graphs \cite{alglave_herding_2014, nienhuis_operational_2016, wickerson_remote-scope_2015, kokologiannakis_effective_2018}, and demonstrating that the axioms hold for each generated partial execution. Some of these works \cite{alglave_herding_2014} performed a machine-checked proof of \circled{I} by showing that the \ism produces exactly the set of all axiomatically consistent outcomes. However, the proof that the executions of the operational model correspond to those of the \ism (\circled{R}) was done by hand, an error-prone approach that has historically missed bugs \cite{manerkar_counterexamples_2016,wickerson_automatically_2017,lahav_repairing_2017}.

In contrast, we machine-check both parts, constructing an \emph{end-to-end} machine-checked proof of axiomatic consistency. The feasibility of machine-checking the entire proof is facilitated by the decoupling of concerns that allows the two proofs to focus on different aspects of correctness in a modular way: the refinement proof (\S\ref{subsec:refinement}) focuses solely on abstracting away the internals of data movement through layers of the caches and the engine running in the operational model, while the proof of axiomatic consistency only reasons about how adding events to the execution graph changes its structure.

\subsection{Connecting the ISM and Operational Model via Refinement}
\label{subsec:refinement}

The goal of the refinement proof is to demonstrate that any execution trace of the operational model (\S\ref{sec:operational}) corresponds to an \ism execution trace that produces partial execution graphs.

What complicates this is the detail in the operational model: most transitions  do not update the execution graph but still meaningfully impact the state for transitions that do. Figure~\ref{fig:proof_anatomy} shows this disconnect in an execution snippet of Figure~\ref{fig:tako_reasoning}a's program. While the operational model performs four transitions ($s1 \rightarrow s5$), only two of them (ending an \onwb{} ($s1 \rightarrow s2$) and starting an \onmiss{} ($s4 \rightarrow s5$)) update the execution graph.
In contrast,
when the \texttt{L3} sends an \onmiss{} request to the \texttt{Network} ($s2 \rightarrow s3$) and the \texttt{Engine} receives it ($s3 \rightarrow s4$), the graph is not updated (as our MCM is at ISA level and does not model unnecessary hardware details).

However, these \textit{internal steps} are still pivotal to correctness: when the \texttt{Engine} later starts an \onmiss{} ($s4 \rightarrow s5$), the address for which it runs and the cache line it populates are determined by the message it receives from the \texttt{Network}, which in turn is determined by the previous internal steps.

Figure \ref{fig:proof_anatomy} shows how refinement reasons about correctness in the presence of internal steps. We first define an abstraction function $abs$ that, for any $s$ of the \tako{} operational model produces an equivalent state in the \ism. Then we show via induction that for any transition $s \rightarrow s'$ of the operational model, $abs(s) \rightarrow abs(s')$ is either a valid \ism transition (e.g., $s1 \rightarrow s2$), or makes no change to the graph (e.g., $s2 \rightarrow s3$).

As we prove refinement through induction,
this involves proving an inductive invariant which shows that the added complexity of caches and \texttt{Network} communication does not change the underlying behavior of the system.

Our proof assumes certain basic properties about our coherence protocol to avoid having to prove coherence in addition to correspondence between our \tako{} implementation model and our ISA-level MCM.
We do so because the coherence protocol we use is well-studied and is known to provide coherence~\cite{nagarajan_primer_2020,oswald_hieragen_2020}.
Additionally, the coherence protocol in a \tako{} system is tangential to \tako{}'s novel features: after phantom data is received from the engine and populates an entry in the directory-level cache, 
the data is indistinguishable from data fetched from memory to the coherence protocol.

We explicitly model all protocol transient states and their interaction with \tako{}, thus verifying any coherence-consistency interface~\cite{manerkar_ccicheck_2015} issues that might arise.
For example, we verify that the dirty bit at directory level is accurately preserved by the coherence protocol, to ensure that \onevict{} and \onwb{} are invoked appropriately.
Even assuming coherence, our proof still required adding 119 clauses to our inductive invariant and 61K LoC of proof annotations.

\section{Related Work}
\label{sec:related}

\textbf{High-Level Language (HLL) and ISA MCMs:}
There is a long line of work that models various ISA and HLL MCMs~\cite{alglave_herding_2014,berghofer_better_2009,adve_weak_1990, batty_mathematizing_2011, pulte_simplifying_2018,manson_java_2005,alglave_gpu_2015,lustig_formal_2019,sarkar_understanding_2011,hutchison_axiomatic_2012}.
Prior work on such models highlighted prefix-closure as a useful property~\cite{kokologiannakis_effective_2018,nienhuis_operational_2016}.
There have been a few papers on MCMs and coherence for novel hardware~\cite{zhang_ila-mcm_2018,ambal_semantics_2024,tan_formalising_2025}, but more research in this area is needed.

\textbf{Hardware Models and Proofs:}
There has also been much work on formally modeling hardware designs using a variety of representations~\cite{burch_automatic_1994,choi_kami_2017,lustig_pipecheck_2014,lustig_coatcheck_2016,subramanyan_template-based_2015,zhang_ila-mcm_2018,wickerson_remote-scope_2015}.
Work on formal verification of hardware implementations includes bounded proofs~\cite{lustig_pipecheck_2014,manerkar_ccicheck_2015,lustig_coatcheck_2016,manerkar_rtlcheck_2017,zhang_coppelia_2018} and complete (all-program) ~\cite{burch_automatic_1994,vijayaraghavan_modular_2015,choi_kami_2017,manerkar_pipeproof_2018,lau_specification_2024,wickerson_remote-scope_2015} proofs.
Some prior hardware verification work uses an intermediate state machine to decompose the proof~\cite{wickerson_remote-scope_2015,alglave_herding_2014}.
However, none of these prior decomposed proofs are completely machine-checked like ours.
Refinement has been used to verify hardware~\cite{burch_automatic_1994, choi_kami_2017, lau_specification_2024} as well as distributed and operating systems~\cite{hawblitzel_ironfleet_2015, lorch_armada_2020, wilcox_verdi_2015, 9617663, gu_certikos_2016, hawblitzel_ironclad_2014, klein_refinement_2010}.
Pensieve~\cite{yang_pensieve_2023} uses uninterpreted functions to overapproximate microarchitectural security behavior, similar to how we overapproximate cache behavior using environmental transitions (\S\ref{subsec:env_trans}).
Our environmental transitions are inspired by
Wickerson et al.~\cite{wickerson_remote-scope_2015}.

\textbf{Programmable Memory Hierarchies:}
Programmable memory hierarchies give software greater control over data movement through the memory hierarchy~\cite{schwedock_tako_2022, kuskin_stanford_1994, mukkara_exploiting_2018, mukkara_phi_2019,schwedock_leviathan_2024}, and can improve performance and energy efficiency.

\section{Conclusion}
\label{sec:conc}

We develop a MCM for the \tako{} PMH, enabling programmers to reason about \tako{} programs without understanding \tako{}'s implementation details.
We also construct a microarchitectural \tako{} model that is parameterized over prefetching policies, cache replacement policies, and network-on-chip specifics.
This model enables architects to change these features in their \tako{} design to improve performance without compromising correctness.
Finally, we prove our MCM sound against our microarchitectural model across all programs, thus verifying that our MCM accurately represents \tako{}.

Our formalization also discovers two more general insights.
First, when creating an ISA-level MCM, ensuring prefix-closure for its axioms makes the MCM amenable to inductive correctness proofs.
Second, microarchitectural models should (and \emph{can}) serve the needs of both architects and formal methods experts, as our microarchitectural model of \tako{} does.

\section*{Acknowledgments}

We thank the anonymous reviewers and our shepherd for their helpful feedback.
We thank Brian Schwedock and Nathan Beckmann for clarifying certain \tako{} implementation details.
This work was supported in part by National Science Foundation grant CCF-2318954.
GitHub Copilot was used for mundane code autocomplete and generation (e.g., find \& replace) in our Dafny proofs. (We handwrote the vast majority of our proofs.)



\section*{Artifact Appendix}

\subsection{Abstract}

This artifact contains two major components:
\begin{enumerate}
\item A Dafny end-to-end machine checked proof of axiomatic consistency for a state machine representation of the täkō hardware (\S\ref{sec:bridge}), which can be explicitly verified.
\item An Alloy encoding of the axioms (Figure~\ref{fig:axioms}) along with the täkō litmus tests presented in the paper, which can be run to confirm the results presented in the paper are consistent with our axioms.
\end{enumerate}

\subsection{Artifact check-list (meta-information)}

{\small
\begin{itemize}
  \item {\bf Run-time environment: } Tested on Windows 11 Enterprise 23H2, with Dafny 4.11.0 and Alloy 6.2.0. Dafny on Windows requires .NET installation. Alloy requires Java (JVM 17+, tested with JVM 24.0.1).
  \item {\bf Hardware: } Tested on 12th Gen Intel(R) Core i7-1280P Windows Laptop with 32 GB RAM.
  \item {\bf Metrics: } Machine-Checked Proof Verification; confirming Litmus Test outcomes for Axiomatic Model.
  \item {\bf Output: } Scripts will output results to the console. Dafny proof expected result: verification script succeeds. Alloy model expected result: Alloy execution matches expected result printed by script.
  \item {\bf Experiments: } Experiment setup and listing included in README file.
  \item {\bf How much disk space required (approximately)?: } With a Dafny installation and Alloy Installation, the full repository size is $\sim$180 MB.
  \item {\bf How much time is needed to prepare workflow (approximately)?: } Around 15 minutes. (installing Dafny + Alloy, Java and potentially .NET).
  \item {\bf How much time is needed to complete experiments (approximately)?:} Around 1.5 hours.
  \item {\bf Publicly available?: } Yes, on GitHub.
  \item {\bf Code licenses (if publicly available)?: } MIT License (included in repository).
  \item {\bf Workflow automation framework used?: } Bash Scripts.
  \item {\bf Archived (provide DOI)?: } Version 1.1.1 uploaded to https://doi.org/10.5281/zenodo.19444275
\end{itemize}
}

\subsection{Description}

\subsubsection{How to access}

The artifact can be cloned from the GitHub Repository at https://github.com/GenericMonkey/takoFormal. The following instructions are all found in the README of this repository as well.

\subsubsection{Hardware dependencies}
The submitted artifact has been tested on an 12th Gen Intel(R) Core(TM) i7-1280P Windows Laptop (32 GB RAM) running Windows 11 Enterprise 23H2. The tools are available for other operating systems and the results should be portable, but there are known brittleness issues with Dafny in particular across different systems. If the scripts are used on a different architecture and certain proofs timeout or face internal errors, please contact.

\subsubsection{Software dependencies}
Installations of Dafny 4.11.0 and Alloy 6.2.0 are required to run the verification scripts. The README contains detailed instructions and links for installation instructions.

\subsection{Installation}

Other than software dependencies mentioned above, no installation is necessary. The files in the repository can be verified using the included bash scripts.

\subsection{Evaluation and expected results}

The two provided scripts correspond to the 2 components of the artifact. Running the \texttt{run\_alloy\_tests.sh} script will iterate through the litmus tests in the repository and confirm the expected outcomes claimed for these tests in the paper. A full table of these is included below, as well as in the README.

\begin{center}
\conf{
\begin{tabular}{|c|c|c|}
\hline
\textbf{Figure} & \textbf{Claimed Result} & \textbf{File} (\texttt{.als}) \\
\hline
Figure \ref{fig:tako_reasoning}a & \makecell[c]{$r1 = 2, r2 = 0$ \\ forbidden} & \texttt{test\_paper\_ex} \\
\hline
Figure \ref{fig:axiom_ex}a & \makecell[c]{$r1 = 1, r2 = 0$ \\ allowed (\S\ref{sec:no_callbacks})} & \texttt{test\_mp} \\
\hline
\makecell[c]{Figure \ref{fig:axiom_ex}a \\ (w/ RMW [b])} & \makecell[c]{$r1 = 1, r2 = 0$ \\ forbidden ($\S$~\ref{sec:no_callbacks})} & \texttt{test\_mp\_rmw} \\
\hline
\makecell[c]{Figure \ref{fig:wbrace}a \\ (w/o i2))} & racy & \texttt{test\_wbrace} \\
\hline
\makecell[c]{Figure \ref{fig:wbrace}a \\ (w/ i2))} & \makecell[l]{a) no race \\ b) $r1 = 0$ \\ \quad forbidden} & \texttt{test\_wbflush} \\
\hline
\makecell[c]{Figure \ref{fig:hats}a \\ (w/o i9))} & racy & \texttt{test\_hatsr} \\
\hline
\makecell[c]{Figure \ref{fig:hats}a \\ (w/ i9))} & \makecell[l]{a) no race \\ b) $r1 \neq r2$ \\ \quad forbidden} & \texttt{test\_hatsnr} \\
\hline
\end{tabular}
}
\ext{
\begin{tabular}{|c|c|c|}
\hline
\textbf{Figure} & \textbf{Claimed Result} & \textbf{File} (\texttt{.als}) \\
\hline
Figure \ref{fig:tako_reasoning}a & \makecell[c]{$r1 = 2, r2 = 0$ \\ forbidden} & \texttt{test\_paper\_ex} \\
\hline
Figure \ref{fig:axiom_ex}a & \makecell[c]{$r1 = 1, r2 = 0$ \\ allowed (\S\ref{sec:no_callbacks})} & \texttt{test\_mp} \\
\hline
\makecell[c]{Figure \ref{fig:mp_extended}a \\ (w/o \onmiss{})} & \makecell[c]{$r1 = 1, r2 = 0$ \\ forbidden} & \texttt{test\_mp\_rmw} \\
\hline
\makecell[c]{Figure \ref{fig:mp_extended}a \\ (w/ \onmiss{})} & \makecell[c]{$r1 = 1, r2 = 0$ \\ forbidden} & \texttt{test\_mp\_rmwcb} \\
\hline
Figure \ref{fig:icbsb}a & \makecell[c]{$r1 = 0, r2 = 0$ \\ allowed} & \texttt{test\_icb\_sb} \\
\hline
\makecell[c]{Figure \ref{fig:wbrace}a \\ (w/o i2))} & racy & \texttt{test\_wbrace} \\
\hline
\makecell[c]{Figure \ref{fig:wbrace}a \\ (w/ i2))} & \makecell[l]{a) no race \\ b) $r1 = 0$ \\ \quad forbidden} & \texttt{test\_wbflush} \\
\hline
\makecell[c]{Figure \ref{fig:phi_extended}a \\ (w/ \onwb{} 1))} & racy & \texttt{test\_phir} \\
\hline
\makecell[c]{Figure \ref{fig:phi_extended}a \\ (w/ \onwb{} 2))} & \makecell[l]{a) no race \\ b) $r1 = 0, r2 =0$ \\ \quad forbidden} & \texttt{test\_phinr} \\
\hline
\makecell[c]{Figure \ref{fig:hats}a \\ (w/o i9))} & racy & \texttt{test\_hatsr} \\
\hline
\makecell[c]{Figure \ref{fig:hats}a \\ (w/ i9))} & \makecell[l]{a) no race \\ b) $r1 \neq r2$ \\ \quad forbidden} & \texttt{test\_hatsnr} \\
\hline
\end{tabular}
}
\end{center}
Running the \texttt{run\_dafny\_verification.sh} script will verify all the Dafny files that make up the end-to-end proof discussed in \S\ref{sec:bridge}. The translation of each axiom presented in Figure~\ref{fig:axioms} to its corresponding file is included in the README.

\subsection{Methodology}

Submission, reviewing and badging methodology:

\begin{itemize}
  \item \url{https://www.acm.org/publications/policies/artifact-review-and-badging-current}
  \item \url{https://cTuning.org/ae}
\end{itemize}

\bibliographystyle{IEEEtranS}
\bibliography{takoformal}

\end{document}